\documentclass[prd,showpacs,superscriptaddress,nofootinbib,floatfix,12pt]{revtex4-2}
\usepackage[english]{babel}
\usepackage[utf8]{inputenc}
\usepackage[utf8]{inputenc}
\usepackage{amsmath,amssymb}
\usepackage{bbold}
\usepackage{slashed}
\usepackage{subfigure}
\usepackage[colorlinks=true,
breaklinks=true,
urlcolor=magenta,
citecolor=blue]{hyperref}
\usepackage[usenames,dvipsnames]{color}
\usepackage{appendix}
\usepackage{braket,bm}
\usepackage{multirow}
\usepackage{enumitem}
\usepackage{color}
\usepackage{cancel}
\usepackage{dsfont}
\usepackage{mathrsfs}
\usepackage{latexsym,bm}
\usepackage{graphicx}
\usepackage{indentfirst}
\usepackage{slashed}
\usepackage{amssymb}
\usepackage{amsmath}
\usepackage{bbm}
\usepackage[utf8]{inputenc}

\graphicspath{{fig/}}

\newcommand{\be}{\begin{equation}} \newcommand{\ee}{\end{equation}}
\newcommand{\ba}{\begin{array}{c}} \newcommand{\ea}{\end{array}}
\newcommand{\bea}{\begin{eqnarray}} \newcommand{\eea}{\end{eqnarray}}

\newcommand{\mS}{{\cal S}}

\newcommand{\mn}{m}
\newcommand{\mpi}{M}
\newcommand{\UVeps}{\epsilon}
\newcommand{\lamb}{\Lambda}

\allowdisplaybreaks

\newcommand{\itp}
{\affiliation{CAS Key Laboratory of Theoretical Physics, Institute of Theoretical Physics, Chinese Academy of Sciences, Beijing 100190, China}}

\newcommand{\phcc}{\affiliation{Peng Huanwu Collaborative Center for Research and Education, Beihang University, Beijing 100191, China}}

\newcommand{\hbnu}{\affiliation{College of Physics and Hebei Key Laboratory of Photophysics Research and Application, Hebei Normal University, Shijiazhuang, Hebei 050024, China}}

\newcommand{\hnu}{\affiliation{School of Physics and Electronics, Hunan University, Changsha 410082, China}}

\newcommand{\huHESPA}{\affiliation{Hunan Provincial Key Laboratory of High-Energy Scale Physics and Applications,\\ Hunan University, Changsha 410082, China}}

\begin{document}
\title{\Large Chiral Representation of the Nucleon Mass at Leading Two-loop Order}

\author{Ze-Rui~Liang}
\email{liangzr@hebtu.edu.cn}
\hbnu\hnu

\author{Han-Xue~Chen}
\email{chenhanxue@hnu.edu.cn}
\hnu

\author{Feng-Kun~Guo}
\email{fkguo@itp.ac.cn}
\itp\phcc

\author{Zhi-Hui~Guo}
\email{zhguo@mail.hebtu.edu.cn}
\hbnu

\author{De-Liang Yao}
\email{yaodeliang@hnu.edu.cn}

\hnu\huHESPA\itp

\begin{abstract}
We calculate the nucleon mass in a manifestly relativistic baryon chiral perturbation theory up to the leading two-loop order. Through dimensional counting analysis, we perform the chiral expansion and verify the validity of the extended-on-mass-shell scheme at the two-loop level. As a result, we obtain the complete chiral representation of the nucleon mass up to $\mathcal{O}(p^5)$, which preserves the original analytic properties and satisfies the correct power counting. The obtained chiral result is well-suited for chiral extrapolation and provides an excellent description of lattice QCD data across a broad range of pion masses. We find that the $\mathcal{O}(p^5)$ contribution is small, approximately $10$~MeV, and varies only mildly with increasing pion mass, demonstrating good convergence of the nucleon mass up to pion masses of about 350~MeV at two-loop order.

\end{abstract}
\date{\today}

\maketitle

\newpage

\section{Introduction}

The nucleon mass, being one of the most fundamental observables in nature, plays a crucial role in our understanding of the origin of visible matter. The decomposition of the proton mass has emerged as a topic of growing interest in recent years,\footnote{Mass decomposition is typically achieved with the help of the energy-momentum tensor (EMT) of quantum chromodynamics. Various kinds of decompositions have been proposed in literature. For instance, the nucleon mass can be defined as the expectation value of the EMT trace, which consequently can be decomposed into two parts: the gluonic trace anomaly and the quark mass term.} primarily motivated by the future Electron-Ion Collider (EIC) in the United States~\cite{Accardi:2012qut} and the Electron-ion collider in China (EicC)~\cite{Anderle:2021wcy}.
Various approaches based on the energy-momentum tensor in quantum chromodynamics (QCD) have been proposed to investigate the decomposition of the proton mass~\cite{Shifman:1978zn,Ji:1994av,Lorce:2017xzd,Metz:2020vxd}, leading to ongoing debates about how the trace anomaly contribution can be further decomposed. Nevertheless, there is consensus that the major part of the nucleon mass comes from  the gluonic part of the trace anomaly, and the rest is from the proton scalar charge, traditionally referred to as the nucleon sigma term.

Given the significance of the nucleon mass, numerous lattice QCD simulations have been performed to study the nucleon mass~\cite{BMW:2008jgk,BMW:2011sbi,Alexandrou:2014sha,Yang:2018nqn,Lin:2019pia,Hu:2024mas}, from which the sigma term can be extracted using the Feynman-Hellmann theorem~\cite{Hellmann:1937book,Feynman:1939zza}. Although calculations at the physical point exist, lattice QCD computations are typically conducted at unphysically large quark masses. Extrapolation to physical quark masses is typically achieved through chiral perturbation theory (ChPT)~\cite{Weinberg:1978kz,Gasser:1983yg,Gasser:1984gg}, a process referred to as chiral extrapolation. The combination of lattice QCD and ChPT has yielded significant advances in our understanding of particle physics~\cite{FlavourLatticeAveragingGroupFLAG:2021npn}.

As a low-energy effective field theory of the Standard Model, ChPT calculations can be systematically improved by including higher-order terms in expansions of quark masses and external momenta (denoted by $p$ in the following)~\cite{Weinberg:1978kz,Gasser:1983yg,Gasser:1984gg}. 
In the purely mesonic sector, many quantities have been calculated beyond one loop in both two- and three-flavor ChPT~\cite{Bellucci:1994eb,Burgi:1996mm,Burgi:1996qi,Bijnens:1997vq,Bijnens:2017wba}. 
Notable examples include two-loop calculations of neutral and charged pion pair photoproduction and their corresponding pion polarizabilities~\cite{Bellucci:1994eb,Burgi:1996mm,Burgi:1996qi}. 
The elastic $\pi\pi$ scattering amplitude has been evaluated to $\mathcal{O}(p^6)$ (two-loop order) in ChPT, along with calculations of pion masses and the decay constant $F_\pi$ to the same order~\cite{Bijnens:1997vq}. 
More recently, three-loop results for masses and decay constants have become available~\cite{Bijnens:2017wba}. For a comprehensive review of ChPT calculations beyond one loop in the mesonic sector, we refer readers to Ref.~\cite{Bijnens:2006zp}.

However, multi-loop calculations involving baryons are more complex than those in the pure-meson sector. The complexity arises from two main factors. First, the emergence of Dirac algebra and the axial coupling between the Goldstone bosons and baryons introduce more Feynman diagram topologies with complicated Lorentz structures and consequently more multiloop master integrals than in the pure mesonic case. Second, the introduction of extra mass scales, specifically the masses of matter fields that are non-zero in the chiral limit, causes the notable power counting breaking (PCB) issue~\cite{Gasser:1987rb}.

As a result, most studies in the baryonic sector are restricted to the one-loop level. A two-loop investigation was first devoted to the nucleon mass~\cite{McGovern:1998tm} in the heavy baryon (HB) formalism, cf. Refs.~\cite{Jenkins:1990jv, Bernard:1995dp}, which found that the HB projection fortunately reduces the number of two-loop master integrals to one. 
The two loops were found to contribute slightly, and the HB chiral series converges very well up to $\mathcal{O}(p^5)$. The nucleon mass was later calculated up to the complete two-loop order~\cite{Schindler:2006ha}, detailed in Ref.~\cite{Schindler:2007dr}, within an alternative renormalization scheme, the infrared regularization (IR) prescription~\cite{Becher:1999he}. 
In fact, the HB result of Ref.~\cite{McGovern:1998tm} can be derived from the IR one obtained in Refs.~\cite{Schindler:2006ha,Schindler:2007dr} by making the heavy baryon expansion. Furthermore, it is found that the so-called leading chiral logarithms are independent of the renormalization scheme, as confirmed by means of the renormalization group equation analysis~\cite{Bijnens:2014lea}.\footnote{The renormalization group equation method has also been applied to the study of the nucleon axial-vector coupling $g_A$ at two-loop order~\cite{Bernard:2006te}.} 
It should be stressed that, as shown in Ref.~\cite{Schindler:2007dr}, a breakdown of the convergence of the $\mathcal{O}(p^6)$ IR result occurs at the unphysical pion mass $\sim 360$~MeV. 
It is known that the IR approach has an unphysical cut~\cite{Bernard:2007zu} due to discarding an infinite series of $M^i$ and $M^i [\ln(M/m)]^j$ ($i$ and $j$ are non-negative integers), where $M$ is the pion mass and $m$ is the nucleon mass in the chiral limit. Thus, the proper analytic property of the renormalized nucleon mass gets spoiled. It would be interesting to investigate whether the convergence improves if the analyticity drawback is removed.

It is acknowledged that the EOMS scheme, proposed in Ref.~\cite{Fuchs:2003qc}, not only has a proper power counting but also has a correct analytic behavior of the physical quantity under consideration. It has been applied to study various important low-energy processes, e.g.~\cite{Yao:2016vbz, GuerreroNavarro:2019fqb, Thurmann:2020mog}, and has been shown to be successful at the one-loop level.
The application of the EOMS scheme at two-loop order is challenging due to the emergence of new PCB terms. For a one-loop calculation, the PCB terms are merely polynomials of small quantities (such as small external momenta and masses of pseudo-Nambu-Goldstone bosons), which can be absorbed by low energy constants (LECs) in the tree-level contributions.
However, for a two-loop computation, non-polynomial PCB terms occur and the treatment of such terms is not trivial at all. The validity of the EOMS scheme at two-loop order was first justified in Ref.~\cite{Schindler:2003je}, albeit in a toy model.
In Ref.~\cite{Djukanovic:2015gna}, it was shown that the application of the complex mass method~\cite{Denner:1999gp}, an analog of the EOMS scheme for unstable particles, to two-loop self-energy diagrams leads to a consistent power counting in a hadronic effective field theory.
The first attempt to carry out a practical application of the EOMS scheme at two-loop order was made in Ref.~\cite{Conrad:2024phd} (see also Ref.~\cite{Conrad:2024sla}) for the nucleon self-energy.
In a recent work~\cite{Chen:2024twu}, an analytical expression of the chiral expansion of the nucleon mass up to $\mathcal{O}(p^6)$ was established by employing the widely used technique of differential equations for Feynman integrals~\cite{Kotikov:1990kg,Kotikov:1991pm, Henn:2013pwa}.
Compared to the IR expression~\cite{Schindler:2006ha,Schindler:2007dr}, only additional terms of $\mathcal{O}(p^4)$ and $\mathcal{O}(p^6)$ enter. These terms are analytic in quark masses and hence have the same structure as the tree-level counterterms of the same chiral orders.
In this regard, the truncated IR and EOMS results are identical since their difference can be compensated by tuning the unknown LECs. The chiral truncation often provides practical convenience, simplifying the result to a concise form as in Eq.~\eqref{eq.ce.p5}. At the physical point, it can usually yield predictions up to a good extent, provided that the chiral series converges well and the contribution of the dropped higher order pieces is negligible.  
However, it destroys the original analytic structure due to the drop of an infinite number of terms, which may be significant for proper chiral extrapolation to a large unphysical pion mass $M$.

In this work, we calculate the nucleon mass using the EOMS scheme up to leading two-loop order, i.e., $\mathcal{O}(p^5)$. 
At this order, all characteristic analytic structures at two-loop level for the nucleon self-energy already appear. 
This can be understood by the fact that the number of master integrals in both the $\mathcal{O}(p^5)$ and $\mathcal{O}(p^6)$ analyses is 13. However, the $\mathcal{O}(p^6)$ calculation introduces more LECs, which are unknown and thus limit the predictive power. The validity of the EOMS scheme at the two-loop level will be demonstrated in detail. 
The analytical expressions of ultraviolet (UV) divergences for the master integrals are explicitly obtained and used to explain the two-loop UV renormalization procedure in baryon ChPT. We find that all the nonlocal UV divergences are exactly canceled by the one-loop subdiagrams, which is a nontrivial check of our two-loop results. The remaining UV divergences are merely monomials in pion masses and are absorbed by the counterterms. 
To extract the PCB terms, we adopt the approach of dimensional counting analysis~\cite{Gegelia:1999qt}, or equivalently the strategy of regions~\cite{Smirnov:2002pj}, which has been used in previous one-loop studies with the EOMS scheme. We show that the PCB terms stemming from regions with at least one hard integration momentum can be removed by shifting the LECs in the tree-level counterterms and one-loop subdiagrams. 
The PCB subtraction procedure is performed in $d$-dimensional spacetime and the relevant $\beta$ functions are obtained. The UV and EOMS renormalization described above results in a two-loop expression of the nucleon mass, which possesses correct power counting, respects the correct analytic properties, and is renormalization scale independent.
By performing expansion of the loop integrals in powers of the pion mass, we can readily reproduce the simplified IR and EOMS results~\cite{Schindler:2006ha,Schindler:2007dr,Chen:2024twu}, up to $\mathcal{O}(p^5)$ within our working accuracy. Note that the calculation of the nucleon mass in Ref.~\cite{Chen:2024twu} was done up to $\mathcal{O}(p^6)$ keeping terms of $M^i (\ln M/\mu)^j$ with $i\le 6$ and $j\le 2$ (c.f. Eq.~\eqref{eq.ce.p5} below). Our full EOMS result is appropriate for chiral extrapolation and can describe lattice QCD data very well for a broad range of pion masses up to at least 350~MeV. 
We also find that the contribution of $\mathcal{O}(p^5)$ is small, $\sim 10$~MeV, and varies mildly as the pion mass increases. This shows good convergence of the chiral representation of the nucleon mass at two-loop order, even up to a large unphysical pion mass $M\sim 300$~MeV.

The outline of the paper is as follows. In Sec.~\ref{sec.se.mn.tlo}, we present the two-loop calculation of the nucleon mass in baryon ChPT. The relevant chiral Lagrangian and definition of the nucleon self-energy are introduced in Sec.~\ref{sec.lag} and Sec.~\ref{sec.se}, respectively. Sec.~\ref{sec.unren.mn} shows the unrenormalized chiral results of the nucleon mass up to $\mathcal{O}(p^5)$. In Sec.~\ref{sec.ren}, renormalization at two-loop order is demonstrated in detail. Subtractions of the UV divergences and PCB terms are discussed in Sec.~\ref{sec.UV.sub} and Sec.~\ref{sec.PCB.sub}, respectively. The full EOMS-renormalized result of the two-loop nucleon mass is presented in Sec.~\ref{sec.full.EOMS.mn}, while its chiral expansion, truncated at $\mathcal{O}(p^5)$, is displayed in Sec.~\ref{sec.chiral.exp}. Numerical results are shown in Sec.~\ref{sec.numerical.result}. A summary and outlook are given in Sec.~\ref{sec.summary}. Definitions, reduction rules, UV divergences and chiral expansion of loop integrals are relegated to Apps.~\ref{sec.appA}, \ref{sec.reduction.rule}, \ref{sec.UV.div} and \ref{sec.dim.counting.analysis}, respectively. Analytical expressions of basic loop integrals needed in the chiral expansion are collected in Appendix~\ref{sec.exp.master.int}. For completeness, the relevant one-loop renormalization factors are compiled in Appendix~\ref{sec.helper}.

\section{Nucleon self-energy and mass at leading two-loop order\label{sec.se.mn.tlo}}

\subsection{Chiral effective Lagrangian\label{sec.lag}}
The chiral effective Lagrangian relevant to our calculation up to the leading two-loop order can be classified as
\begin{align}
\mathcal{L}_{\rm eff}=
\mathcal{L}^{(2)}_{\pi \pi}+\mathcal{L}^{(4)}_{\pi \pi}+
\mathcal{L}^{(1)}_{\pi N}+\mathcal{L}^{(2)}_{\pi N}+\mathcal{L}^{(3)}_{\pi N}+\mathcal{L}^{(4)}_{\pi N} \ ,
\end{align}
with the superscripts denoting the chiral orders. Specifically, the purely mesonic parts read~\cite{Gasser:1983yg}
\begin{align}
\mathcal{L}^{(2)}_{\pi\pi}&=\frac{F^2}{4}\left\langle
u_{\mu}u^{\mu}+\chi_+
\right\rangle \ , \\
\mathcal{L}^{(4)}_{\pi \pi}&=\frac{1}{8}\ell_4\left\langle
u_{\mu}u^{\mu}\right\rangle\left\langle\chi_+
\right\rangle 
+\frac{1}{16}(\ell_3+\ell_4)\left\langle\chi_+
\right\rangle^2
\ .
\end{align}
where $\langle\cdots\rangle$ represents the trace in the flavor space, $F$ is the pion decay constant in the chiral limit, and $\ell_3$, $\ell_4$ are mesonic LECs. The building blocks $u_\mu$ and $\chi_\pm$ are given by 
\begin{align}
u_\mu&=i\left[u^\dagger \partial_\mu u-u\partial_\mu u^\dagger \right]
\ , \qquad u=\exp\left[\frac{i\pi_a \tau_a}{2F_0}\right] ,
\\
\chi_\pm&=u^\dagger \chi u^\dagger \pm u\chi^\dagger u \ , \qquad \chi={\rm diag}\{M^2, M^2\} \ , 
\end{align}
with $\pi_a$ ($a=1,2,3$) being the pion fields and $\tau_a$ being the usual Pauli matrices. Here $M^2$ is the pion mass squared at the lowest order. It is related to the light quark masses via the Gell-Mann--Oakes--Renner relation
\begin{align}
    M^2=2B \hat{m}\ ,\quad \hat{m}=(m_u+m_d)/2\ ,
\end{align}
where the constant $B$ is proportional to the quark condensate~\cite{Scherer:2012xha}. Note that we work in the limit of exact isospin symmetry, i.e., $m_u=m_d$.

The Lagrangian describing pion-nucleon interactions reads~\cite{Fettes:2000gb}
\begin{align}
\mathcal{L}^{(1)}_{\pi N}&=\bar{N}\left\{
i\slashed{D}-m-\frac{1}{2}g\slashed{u}\gamma^5
\right\}N\ , \label{eq.lagrangian.lo.piN} \\
\mathcal{L}^{(2)}_{\pi N}&=\bar{N}\left\{
c_1\left\langle\chi_+\right\rangle
-\frac{c_2}{4m^2}\left\langle u_{\mu}u^{\mu}\right\rangle(D_{\mu}D_{\nu}+h.c.)
+\frac{c_3}{2}\left\langle u_{\mu}u^{\mu}\right\rangle 
-\frac{c_4}{4}\gamma^{\mu}\gamma^{\nu}\left[u_{\mu},u_{\nu}\right]
\right\} N \ ,\\
\mathcal{L}^{(3)}_{\pi N}&=\bar{N}\left\{
\frac{d_{16}}{2}\gamma^{\mu}\gamma^5\left\langle\chi_+\right\rangle u_{\mu}
+i\frac{d_{18}}{2}\gamma^{\mu}\gamma^5\left[D_{\mu},\chi_{-}\right]
\right\} N \ ,\\
\mathcal{L}^{(4)}_{\pi N}&=\bar{N}\bigg\{
e_{38}\left\langle\chi_+\right\rangle\left\langle\chi_+\right\rangle
+e_{39}\tilde{\chi}_+\left\langle\chi_+\right\rangle
+e_{40}\left\langle\tilde{\chi}_+ \tilde{\chi}_+\right\rangle
+\frac{e_{115}}{4}\left\langle \chi_+^2-\chi_-^2\right\rangle 
\notag \\
&\hspace{1.0cm}
-\frac{e_{116}}{4}\left(
\left\langle \chi_-^2\right\rangle 
-\left\langle \chi_-\right\rangle^2
+\left\langle \chi_+^2\right\rangle 
-\left\langle \chi_+\right\rangle^2
\right)
\bigg\} N \ ,
\end{align}
where $m$ and $g$ denote the nucleon mass and the axial-vector coupling constant in the chiral limit, respectively. The higher-order LECs $c_{i=1,2,3,4}$, $d_{j=16,18}$ and $e_{k=38,39,40,115,116}$ are in units of GeV$^{-1}$, GeV$^{-2}$ and GeV$^{-3}$, in order. The covariant derivative $D_\mu$ acting on the nucleon field $N$ is defined by
\begin{align}
D_\mu N=(\partial_\mu+\Gamma_\mu)N \ , 
\quad \Gamma_\mu=\frac12\left[u^\dagger\partial_\mu u+u\partial_\mu u^\dagger\right]\ ,\quad
N=\left(\begin{array}{c}
     p  \\
     n 
\end{array}
\right) ,
\end{align}
where $\Gamma_\mu$ is the chiral connection. The building block $\tilde{\chi}_+$ is traceless and given by
\begin{align}
\tilde{\chi}_+=\chi_+-\frac12\left\langle \chi_+\right\rangle \ .
\end{align}
Note that very recently, the pion-nucleon Lagrangian has been constructed up to $\mathcal{O}(p^5)$ in Ref.~\cite{Song:2024fae}, although it is irrelevant to our current calculation. That is, the LECs from the $\mathcal{O}(p^5)$ Lagrangian do not enter the calculation of the nucleon mass up to the same order.

\subsection{General structure of the nucleon self-energy up to the leading two-loop order\label{sec.se}}

The dressed propagator of the nucleon is given by 
\begin{align}
i S_N(p)=\frac{i}{\slashed{p}-m-\Sigma_N(\slashed{p})}=\frac{i{\cal Z}_N}{\slashed{p}-m_N}+\text{non-pole~piece}\ ,
\end{align}
where $-i\Sigma_N$ denotes the nucleon self-energy and is a sum of the one-particle irreducible diagrams contributing to the nucleon two-point function. The physical nucleon mass $m_N$ is defined as the pole of the dressed propagator, while the residue $\mathcal{Z}_N$ of the pole term corresponds to the wave function renormalization constant. 

\begin{figure}[tb]
\centering
\includegraphics[width=0.5\textwidth]{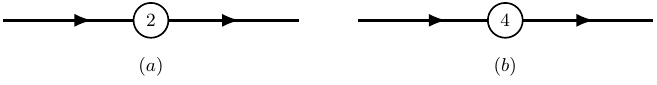}
\caption{Tree-level Feynman diagrams contributing to the nucleon self energy, where the circled numbers denote the chiral orders of the insertions.}
\label{fig:mn_tree}
\end{figure}

For a calculation up to and including $\mathcal{O}(p^5)$, the nucleon self-energy comprises tree, one-loop and two-loop diagrams shown in Figs.~\ref{fig:mn_tree}, \ref{fig:1loop} and~\ref{fig:2loop}, respectively. The corresponding expression can be written in the form
\begin{align}
\Sigma_N(\slashed{p},m)=-4c_1 M^2  -2 e_m M^4 +\hbar\, \Sigma^{(1)}_N(\slashed{p},m_4)+\hbar^2\, \Sigma^{(2)}_N(\slashed{p},m_4)
\end{align}
with ${e}_m=e_{115}+e_{116}+8e_{38}$. The reduced Planck constant $\hbar$ is used as a counting parameter, whose power indicates the number of loops. The first two terms are tree-level contributions of $\mathcal{O}(p^2)$ and $\mathcal{O}(p^4)$, respectively. 
Note that the second variable in the one- and two-loop self-energies is $m_4=m-4c_1M^2-2e_m M^4$, such that the contributions with mass insertions on the internal nucleon lines are automatically incorporated. 
The nucleon mass can be obtained by solving the equation
\begin{align}
\big[\slashed{p}-m_4-\Sigma_N(\slashed{p},m_4)\big]_{\slashed{p}=m_N}=0\ .
\end{align}
At leading two-loop order, it can be recast as
\begin{align}
m_N-m_4-\hbar\Sigma_N^{(1)}(m_N,m_4)-
\hbar^2\Sigma_N^{(2)}(m_N,m_4)=0\ .\label{eq.se.eq}
\end{align}
The solution has the form
\begin{align}\label{eq.mnsol}
m_N=m_4 + \hbar \Delta m_N^{(1)}  + \hbar^2 \Delta m_N^{(2)}\ . 
\end{align}
Inserting it into the right-hand side of Eq.~\eqref{eq.se.eq} and expanding up to order $\hbar^2$, one obtains
\begin{align}
\Delta m_N^{(1)} &=  \Sigma_N^{(1)}(m_4,m_4) \ ,\\
\Delta m_N^{(2)} &= \Sigma_N^{(1)}(m_4,m_4)\Sigma_N^{(1)\prime}(m_4,m_4)+ \Sigma_N^{(2)}(m_4,m_4)\ .
\end{align}
The derivative of the one-loop self-energy is related to the corresponding one-loop wave function renormalization constant via
\begin{align}
\Sigma_N^{(1)\prime}(m_4,m_4)
\equiv \left[\frac{\partial  }{\partial \slashed{p}}\Sigma_N^{(1)}(\slashed{p},m_4)\right]_{\slashed{p}=m_4}=\mathcal{Z}_N^{(\text{1-loop})}-1\equiv\delta \mathcal{Z}_N^{(\text{1-loop})}\ .
\end{align}
A naive chiral expansion yields
\begin{align}
\Sigma_N^{(1)}(m_4,m_4)\Sigma_N^{(1)\prime}(m_4,m_4)=
m_{N}^{(1a)}\,\delta \mathcal{Z}_N^{(2)}+\cdots,
\label{eq:1loto2lo}
\end{align}
where the ellipsis represents higher-order pieces beyond $\mathcal{O}(p^5)$. Explicit expressions for $m_{N}^{(1a)}$ and $\delta\mathcal{Z}_N^{(2)}$ are given by Eqs.~\eqref{eq.mn.1a.1loop} and~\eqref{eq.1loop.ZN}, respectively.

\subsection{Unrenormalized result\label{sec.unren.mn}}

\begin{figure}[h]
\centering
\includegraphics[width=\textwidth]{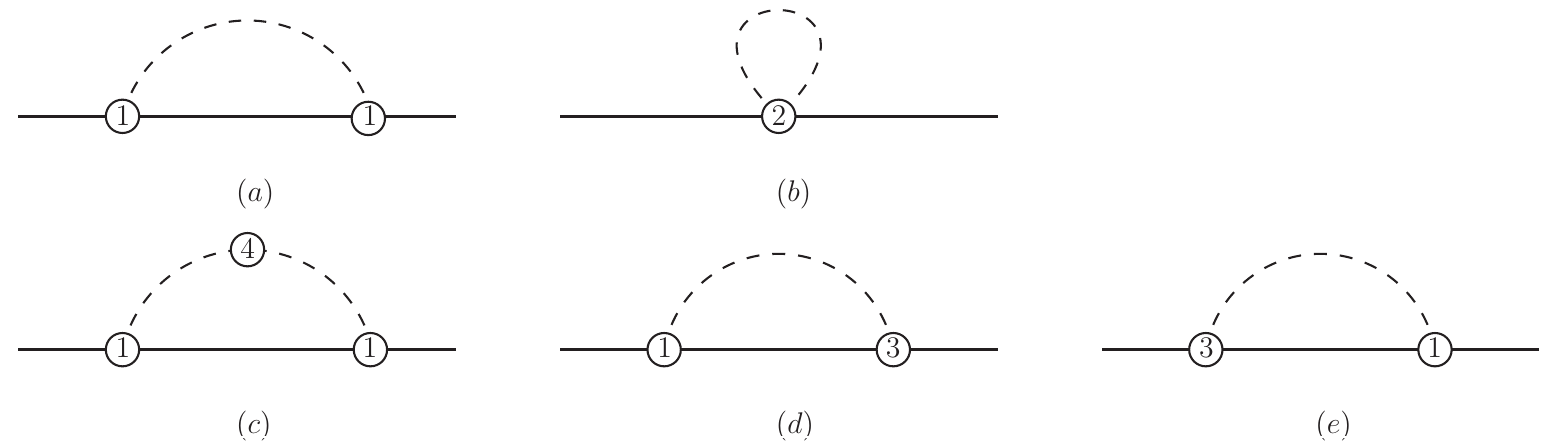}
\caption{One-loop Feynman diagrams contributing to the nucleon self-energy. The solid and dashed lines represent nucleons and pions, respectively. The circled numbers indicate the chiral dimensions of the vertices.}
\label{fig:1loop}
\end{figure}

In this subsection, we calculate the last two terms in Eq.~\eqref{eq.mnsol}. 
To obtain $\Delta m_N^{(1)}$, we need to calculate the one-loop self-energy Feynman diagrams shown in Fig.~\ref{fig:1loop}. Note that the nucleon propagates with the auxiliary mass $m_4$ in the loop diagrams.

The one-loop contributions up to $\mathcal{O}(p^5)$ are calculated diagram by diagram in Fig.~\ref{fig:1loop} and the sum is
\begin{align}
   \Delta m_N^{(1)} = m_N^{(1a)} + m_N^{(1b)} + m_N^{(1c)} + m_N^{(1d)} + m_N^{(1e)}\ ,
\end{align}
where the explicit expression for each individual term reads 
\begin{align}
\mathcal{O}(p^3):\quad   
m_{N}^{(1a)}&=\frac{3g^2m\,\kappa}{2{i}F^2}\left[ 
J_{01}+M^2 J_{11} 
\right]\ ,\label{eq.mn.1a.1loop}\\
\mathcal{O}(p^4):\quad    
m_{N}^{(1b)}&=\frac{3M^2\,\kappa}{idF^2}\left[c_2+(c_3-2c_1)d\right] J_{10}\,
\ ,\label{eq.mn.1b}\\
\mathcal{O}(p^5):\quad  {
m_{N}^{(1c)}}&=-\frac{3g^2 m\,\kappa}{2i F^4(4m^2-M^2) }\bigg\{
2 J_{01} \bigg[(d-2) \ell_3 M^4+\ell_4 \left(4 M^2 m^2-M^4\right)\bigg]+M^4 \\
&
\times\bigg[2 J_{11} \left((d-2) \ell_3 M^2-2 (d-1) \ell_3 m^2+\ell_4 \left(4 m^2-M^2\right)\right)-(d-2) \ell_3 J_{10}\bigg]
\bigg\}\notag
,\\
\mathcal{O}(p^5):\quad    m_{N}^{(1d)}&=m_{N}^{(1e)}=
\frac{3g(2d_{16}-d_{18})m M^2\,\kappa}{i F^2}
\left[ 
J_{01}+M^2 J_{11}
\right]
\ .
\end{align}
Again, the quantity $m$ in the above expressions should be regarded as $m_4$. Here, $d$ denotes the dimension of spacetime. The definitions of the one-loop integrals $J_{01,10,11}$ and the overall factor $\kappa$ are specified by Eq.~\eqref{eq.def.olo.int} and Eq.~\eqref{eq.kappa} in Appendix~\ref{sec.appA}, respectively. To make the one-loop results more compact, we have already carried out the so-called Passarino-Veltman (PV) reductions~\cite{Passarino:1978jh} for the one-loop integrals. It is noticed that, at one-loop order, all the loop integrals are reduced to the linear combinations of three master integrals: 
\begin{align}
\label{eq.olo.master.int}
 J_\text{master}^\text{1-loop} =  \{J_{10},\, J_{01},\, J_{11}\}\ .
\end{align}

\begin{figure}[ht]
\centering
\includegraphics[width=\textwidth]{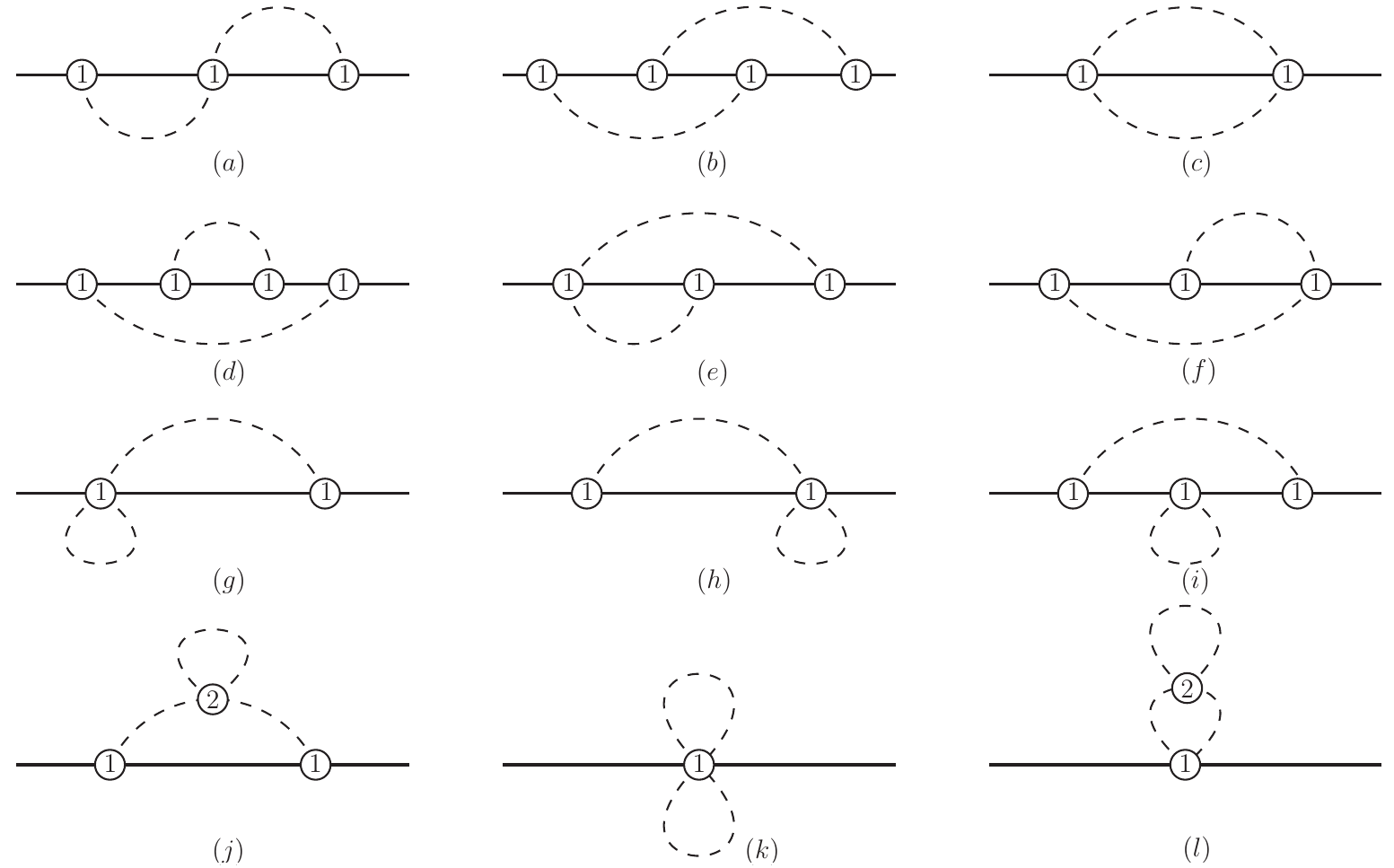}
\caption{Two-loop Feynman diagrams contributing to the nucleon self-energy. The solid and dashed lines represent nucleons and pions, respectively. The circled numbers indicate the chiral dimensions of the vertices.}
\label{fig:2loop}
\end{figure}

\begin{table}[ht]
\centering
\caption{Symmetric factors $\mS$ for the two-loop diagrams in Fig.~\ref{fig:2loop}.}
\label{tab:sf}
\begin{ruledtabular}
\begin{tabular}{l|ccccccccccccc}
Diagram& (a) & (b)& (c) & (d) & (e) & (f) & (g) &(h) & (i) & (j) &(k)&(l)  \\
\hline
$\mS$  & 1 & 1 & 2 & 1 & 1 &1 & 2 & 2 & 2 & 2 & 4 & 4 \\
\end{tabular}
\end{ruledtabular}
\end{table}

Likewise, for $\Delta m_N^{(2)}$, the Feynman diagrams of two loops that contribute to the self-energy of the nucleon are depicted in Fig.~\ref{fig:2loop}.
Symmetric factors $\mS$ for the two-loop diagrams are all collected in Table~\ref{tab:sf}. The explicit forms corresponding to the unrenormalized two-loop contributions from the diagrams in Fig.~\ref{fig:2loop} are given by  
\begin{itemize}
\item Diagram (a):
\begin{align}
  m_{N}^{(2a)} &= \frac{3g^2\,\kappa^2}{16 m F^4\mS}\bigg[8m^2 M^4 I_{11011} + 8m^2 M^2 (I_{01011}+I_{10011}) +2m^2 \big[4I_{00011}
  \notag\\
  &+  I_{11(-1)01} 
  + I_{11(-1)10}  - I_{110(-1)1} - I_{1101(-1)}\big] +2(M^2-2m^2)I_{11000} \notag \\
  &
  -I_{110(-1)0}-I_{1100(-1)}
  \bigg] ,
\end{align}
\item Diagram (b):
\begin{align}
  m_N^{(2b)}&= -\frac{3 g^4\,\kappa^2}{32 F^4 m\mS}  \bigg[16 M^4 m^4 I_{11111}+16M^2 m^4
   I_{01111} +16M^2
   m^4 I_{10111} +4M^4 m^2 I_{11011} \notag\\
 &+4 M^4 m^2 I_{11101} +4M^4 m^2 I_{11110} +4M^2 m^2 I_{01011}
   +4M^2 m^2 I_{01101} \notag\\
 &+4M^2 m^2
   I_{01110} +4 M^2
   m^2 I_{10011}+4M^2 m^2 I_{10101} +4M^2 m^2 I_{10110} \notag\\
 &-4 M^2 m^2 I_{11001} -4M^2 m^2 I_{11010}
   +8M^2 m^2 I_{11100} +16m^4
   I_{00111} \notag\\
 &-8 m^4 I_{11(-1)11}+16m^4 I_{11001} +16 m^4
   I_{11010}-8 m^4 I_{111(-1)1}-16m^4 I_{11100} \notag\\
 &-8m^4
   I_{1111(-1)} +4 m^2I_{00011} +4m^2 I_{00101} +4m^2
   I_{00110} -4 m^2 I_{01010} \notag\\
 &+4 m^2 I_{01100}-4m^2
   I_{10001} +4 m^2 I_{10100} +8m^2 I_{11000} -8m^2
   I_{111(-1)0} \notag\\
 &-8 m^2 I_{1110(-1)} +M^4 I_{11100} +M^2 I_{01100} + M^2I_{10100}-I_{111(-1)(-1)}\bigg] ,
\end{align}
\item Diagram (c):
\begin{align}
  m_N^{(2c)}&=-\frac{3\,\kappa^2}{16m F^4\mS}\bigg[
  2M^2\left(2 M^2 I_{11100}+2I_{01100}+2I_{10100}-I_{11000}\right)
  \notag\\
  &
  +I_{110(-1)0}+I_{1100(-1)}-4I_{111(-1)(-1)}
  \bigg]
  \label{eq.mn.2c} ,
\end{align}
\item Diagram (d):
\begin{align}
 m_N^{(2d)}&= -\frac{9 g^4\,\kappa^2}{3
   2 F^4 m\mS }\bigg[16M^4 m^4I_{11120} +16M^2 m^4
   I_{01120} +16 M^2
   m^4 I_{10120}+4 M^4 m^2 I_{11020} \notag\\
  &+8 M^4 m^2 I_{11110} +4 M^2 m^2 I_{01020} +8M^2 m^2 I_{01110} 
   \notag\\
  &+8M^2 m^2
   I_{10110} -4M^2
   m^2 I_{11(-1)20} +8 M^2 m^2 I_{11010} -8M^2 m^2 I_{11100} \notag\\
  &+16 m^4I_{00120} 
  +8m^2
   I_{00110} -4m^2 I_{01(-1)20} +8m^2 I_{01010} \notag\\
  &-4m^2
   I_{01100} -4 m^2 I_{10100} +M^4 I_{11100} +M^2 I_{01100} +M^2 I_{10100} -M^2I_{11000}
   \notag\\
  &
  +I_{1100(-1)}-I_{111(-1)(-1)}\bigg] ,
\end{align}
\item Diagram (e) and (f):
\begin{align}
 m_N^{(2e)}= m_N^{(2f)}
 &=\frac{3 g^2\,\kappa^2}{16
   F^4 m\mS} \bigg[8 M^4 m^2I_{11101} +8 M^2 m^2
   I_{01101}+8M^2
   m^2 I_{10101} +8 m^2 I_{00101}\notag\\
 &-2 m^2 I_{11(-1)01}+2m^2 I_{110(-1)1}
   +6 m^2 I_{11000}-4 m^2 I_{111(-1)0}\notag\\
 &-4m^2 I_{1110(-1)}
   +2  M^4I_{11100}+2M^2 I_{01100} +2 M^2
   I_{10100}\notag\\
 &-M^2I_{11000} 
 +I_{110(-1)0}-2 I_{111(-1)(-1)}\bigg] ,
\end{align}
\item Diagram (g) and (h): 
\begin{align}
 m_N^{(2g)}&= m_N^{(2h)}=-\frac{g^2\,\kappa^2}{4m F^4\mS}\bigg[
 4 M^2 m^2I_{11001} +
 4 m^2I_{10001} +M^2I_{11000} 
 -I_{1100(-1)} \bigg] ,
\end{align}
\item Diagram (j):
\begin{align}
 m_N^{(2j)}&= 
 \frac{g^2\,\kappa^2}{8 F^4 m\mS} \bigg[12 M^4 m^2I_{12001} +28M^2
   m^2 I_{11001}  
   +16 m^2I_{10001} 
   +3 M^4 I_{12000} \notag\\
   &+7M^2 I_{11000} -3M^2 I_{1200(-1)} 
   -4 I_{1100(-1)}\bigg]
 \ .
\end{align}
\end{itemize}
The remaining Feynman diagrams in Fig.~\ref{fig:2loop} do not contribute, i.e., $m_N^{(2i)}= m_N^{(2k)}=m_N^{(2l)} =0$. 
The reason is that the leading order $NN\pi\pi$ vertex is from a vector $\pi\pi$ pair from the Weinberg-Tomozawa term as can be seen from Eq.~\eqref{eq.lagrangian.lo.piN}, however, a nonvanishing tadpole contribution to the nucleon mass requires a scalar $\pi\pi$ pair. Therefore, any pion tadpole diagram with the leading order $NN\pi\pi$ vertex must vanish.

The term in Eq.~\eqref{eq:1loto2lo} also contributes at the two-loop order and is particularly important for the renormalization to be discussed in Sec.~\ref{sec.ren}. The two-loop results can be simplified using the integration-by-parts (IBP) reduction~\cite{Smirnov:2019qkx}, with the relevant reduction rules compiled in Appendix~\ref{sec.reduction.rule} for completeness. After reduction, the two-loop contributions are expressed in terms of $26$ master integrals, of which $9$ are separable and can be written as products of one-loop master integrals in Eq.~\eqref{eq.olo.master.int}. Furthermore, the number of independent master integrals reduces to $10$ when applying the symmetry of interchanging $\ell_1$ and $\ell_2$ in the first two propagators and in the last two, respectively, in Eq.~\eqref{eq.int.tlo.def}. The minimal set of independent two-loop master integrals reads
\begin{align}\label{eq.tlo.master.int}
I_\text{master}^\text{2-loop} =\{  & I_{11111},\, I_{11110}\leftrightarrow I_{11101},\, I_{10111}\leftrightarrow I_{01111},\, I_{21100}\leftrightarrow I_{12100},\,
I_{11100},\, I_{00111},\, \notag\\
&I_{10101}\leftrightarrow I_{01110},\, I_{10110}\leftrightarrow I_{01101},\, I_{10100}\leftrightarrow I_{01100},\,
I_{00110}\leftrightarrow I_{00101}\}\ .
\end{align}
Note that the dot-basis of master integrals is chosen in the reduction procedure known as the Laporta algorithm~\cite{Laporta:2000dsw}. This basis avoids irreducible scalar products, meaning no negative $\nu_{i}$ ($i=1,\cdots,5$) indices appear, though at the expense of introducing positive indices larger than one (see $I_{21100}$ in Eq.~\eqref{eq.tlo.master.int}). The choice of dot-basis is appropriate for carrying out EOMS renormalization using the method of dimensional counting analysis, which will be discussed in the next section.

\section{Renormalization\label{sec.ren}}

The applicability of EOMS and its generalization, the complex mass scheme, at two-loop order was first demonstrated in Refs.~\cite{Schindler:2003je,Djukanovic:2015gna}, respectively, but using toy-model Lagrangians. In this section, we present a detailed illustration of the validity of the EOMS scheme at the two-loop level using the realistic example of the nucleon mass in ChPT.

\subsection{UV subtraction\label{sec.UV.sub}}
\subsubsection{Special case: $g=0$}

For ease of explanation, we first consider a simple case by taking the axial-vector coupling $g=0$, i.e., removing the $\pi NN$ interaction vertex. With $g=0$, only the two tree diagrams of Fig.~\ref{fig:mn_tree}, one-loop diagram (b) of Fig.~\ref{fig:1loop} and two-loop diagram (c) of Fig.~\ref{fig:2loop} are involved. The $\mathcal{O}(p^5)$ $m_N$ simplifies to
\begin{align}
m_N = m-4c_1 M^2-2e_m M^4 +m_N^{(1b)}+m_N^{(2c)}\ ,\label{eq.mN.gAeq0}
\end{align}
where the unrenormalized expressions of $m_N^{(1b)}$ and $m_N^{(2c)}$ are given by Eq.~\eqref{eq.mn.1b} and Eq.~\eqref{eq.mn.2c}, respectively.

The expressions of the UV divergences of master integrals are explicitly given in Appendix~\ref{sec.UV.div}. Then, the UV $\epsilon$-pole terms of $m_N^{(2c)}$ can be obtained, which are
\begin{align}
 {\rm UV}[m_N^{(2c)}]&=\frac{3 \left(6 \mpi^4 \mn+8 \mpi^2 \mn^3-3 \mn^5\right)}{32 F^4 \lamb^2 \UVeps^2}+\frac{132 \mpi^4 \mn-16 \mpi^2 \mn^3-9 \mn^5}{128 F^4 \lamb^2 \UVeps}
 \notag\\
 &+\frac{3 \mn^3 \left(3 \mn^2-8 \mpi^2\right) }{16 F^4 \lamb^2 \UVeps}\log \frac{\mn^2}{\mu^2}-\frac{9 \mpi^4 \mn }{8 F^4 \lamb^2 \UVeps}\log \frac{\mpi^2}{\mu^2}\ ,\label{eq.mN2c.UV}
\end{align}
with $\Lambda=1/(16\pi^2)$. The last term on the right-hand side, accompanied by the logarithm of pion mass, corresponds to the so-called non-local sub-divergence, which cannot be absorbed by the tree-level counter terms. Nevertheless, it can be canceled by one-loop diagrams with insertions from the next-to-leading order vertices, for instance, the one-loop diagram in Fig.~\ref{fig:1loop}~(b). 

To be specific, the unrenormalized form of the one-loop diagram $(b)$ reads
\begin{align}
  m_N^{(1b)} = -\frac{3    (d (c_3-2 c_1)+c_2)\mpi^4}{F^2 d}\bigg[
  -\frac{1}{\lamb \UVeps}
  +\frac{ 1}{\lamb}\log\frac{\mpi^2}{\mu^2}+\mathcal{O}(\epsilon)\bigg]\ .\label{eq.UV.1lo.1b}
\end{align}
It should be emphasized that the finite piece must be kept. To proceed, the bare LECs, {i.e.} $c_1$, $c_2$ and $c_3$, are split as
\begin{align}
X=-\frac{\beta_X^{(1)}}{\epsilon\lamb}+X^r \ ,\quad X\in\{c_1,c_2,c_3\}\ .\label{eq.olo.UV.gAeq0}
\end{align}
Here, the one-loop beta functions of $c_{1,2,3}$ are taken from the analysis of pion-nucleon scattering in Ref.~\cite{Yao:2016vbz}:
\begin{align}
\beta_{c_1}^{(1)}=0 \ ,\quad \beta_{c_2}^{(1)}=\frac{ m}{2F^2}\ ,\quad \beta_{c_3}^{(1)}=\frac{m}{4F^2} \ ,
\end{align}
and only the pieces surviving under the $g=0$ condition are retained. Substituting Eq.~\eqref{eq.olo.UV.gAeq0} into Eq.~\eqref{eq.UV.1lo.1b}, we get the UV divergence part
\begin{align}
   {\rm UV}[m_N^{(1b)}] =-\frac{9 \mpi^4 \mn}{8 F^4 \lamb^2 \UVeps^2}
   -\frac{3 \mpi^4 \mn}{16 F^4 \lamb^2\UVeps}+\frac{9 \mpi^4 \mn}{8 F^4 \lamb^2\UVeps}\log\frac{M^2}{\mu^2}
   -\frac{3 \mpi^4 (8 c_1^r-c_2^r-4 c_3^r)}{4 F^2 \lamb\UVeps}\ .\label{eq.mN1b.UV}
\end{align}
The terms proportional to $\log[M^2/\mu^2]$ in Eq.~\eqref{eq.mN2c.UV} and Eq.~\eqref{eq.mN1b.UV} are opposite in sign, and hence they cancel each other out. The sum of the one- and two-loop UV divergences leads to
\begin{align}
{\rm UV}[m_N^{(1b)}+m_N^{(2c)}] &=\frac{3 \left(-6 \mpi^4 \mn+8 \mpi^2 \mn^3-3 \mn^5\right)}{32 F^4 \lamb^2 \UVeps^2}+\frac{108 \mpi^4 \mn-16 \mpi^2 \mn^3-9 \mn^5}{128 F^4 \lamb^2 \UVeps}\notag\\
&-\frac{3 \mpi^4 (8 c_1^r-c_2^r-4 c_3^r)}{4 F^2 \lamb\UVeps}\ .
\end{align}
The net UV divergence contains only polynomials in the pion mass $M$ and can be absorbed by the bare parameters in the tree-level counter terms. That is, the quantities $m$, $c_1$, and $e_m$ appearing in Eq.~\eqref{eq.mN.gAeq0} can be split as
\begin{align}\label{eq.LECs.two.loop}
X=\frac{\beta_X^{(22)}}{\epsilon^2\Lambda^2}
-\frac{\beta_X^{(21)}}{\epsilon\Lambda^2}-\frac{\beta_X^{(11)}}{\epsilon\Lambda}+X^r\ ,\quad X\in\{m, c_1, e_m\}\ ,
\end{align}
where the first and second superscripts, $i$ and $j$, of the beta function $\beta^{(ij)}_X$ stand for the number of loops and the power of UV divergence $-1/\epsilon$, respectively. Finally, we obtain the beta functions for $m$, $c_1$ and $e_m$ at two-loop order with $g=0$:
\begin{align}
&\beta_{m}^{(22)} =\frac{9m^5}{32 F^4}\ ,\quad 
\beta_{m}^{(21)} =-\frac{9m^5}{128 F^4}\ ,\quad
\beta_{m}^{(11)} = 0\ ,\notag\\
&\beta_{c_1}^{(22)} =\frac{3m^3}{16 F^4}\ ,\quad 
\beta_{c_1}^{(21)} =\frac{m^3}{32 F^4}\ ,\quad 
\beta_{c_1}^{(11)} = 0\ ,\notag\\
&\beta_{e_m}^{(22)} =-\frac{9m}{32 F^4}\ ,\quad 
\beta_{e_m}^{(21)} =-\frac{27m}{64 F^4}\ ,\quad 
\beta_{e_m}^{(11)} = \frac{3(8c_1-c_2-4c_3)}{8F^2}\ .\label{eq.beta.twoloop.UV}
\end{align}

\subsubsection{General case: $g\neq 0$}

The above procedure can be readily generalized to the realistic case with $g\neq 0$, although much more complicated and tedious. Two-loop diagrams contain one-loop ones as their subdiagrams.
More LECs are now involved compared to the $g=0$ case. As a consequence, Eq.~\eqref{eq.olo.UV.gAeq0} is extended to
\begin{align}
 X=-\frac{\beta_X^{(1)}}{\epsilon\lamb}+X^r \ ,\quad X\in \{m,g,c_1,c_2,c_3,\ell_3,\ell_4,d_{16},d_{18}\}\ ,
 \label{eq45}
\end{align}
with the one-loop $\beta$ functions given by~\cite{Chen:2012nx,Siemens:2016hdi,Gasser:1983yg}
\begin{align}
&\beta_m^{(1)}=\frac{3g^2 m^3}{2 F^2  }\ ,\quad
\beta_g^{(1)}=\frac{g(-2+g^2)m^2}{F^2}\ ,\quad
\beta_{c_1}^{(1)}=-\frac{3g^2 m}{8 F^2  }\ ,\notag\\ 
&\beta_{c_2}^{(1)}=\frac{(g^2-1)^2 m}{2F^2} \ , \quad
\beta_{c_3}^{(1)}=\frac{(1-6g^2+g^4)m}{4F^2}\ ,\quad
\beta_{\ell_3}^{(1)}=- \frac{1}{4}\ , \notag\\
&\beta_{\ell_4}^{(1)}=1\ ,\quad \beta_{d_{16}} = \frac{g(-1 + g^2)}{4 F^2 }\ ,\quad  \beta_{d_{18}} = 0\ .\label{eq.EOMS.UV.beta}
\end{align}
We have verified that the non-local subdivergences, i.e., the terms proportional to $(1/\epsilon)\ln M$, from the two loops are exactly canceled out by the one-loop diagrams with a vertex from the next-to-leading order operators if the above $\beta$ functions are used. This offers a nontrivial consistency check of the correctness of our results.

In addition, the remaining UV divergences can be absorbed by the LECs from the counterterms in the way specified in Eq.~\eqref{eq.LECs.two.loop}. However, the two-loop $\beta$ functions in Eq.~\eqref{eq.beta.twoloop.UV} become more complex due to the non-zero $g$ coupling. To remove the two-loop UV divergences of the types $1/\epsilon^2$ and $1/\epsilon$, we obtain
\begin{align}
\beta_{m}^{(22)} &= \frac{9 \left(5 g^4-2 g^2+1\right) \mn^5}{32 F^4}\ ,\quad \beta_m^{(21)} = \frac{\left(223 g^4-14 g^2-9\right) \mn^5}{128 F^4}\ ,\\
\beta_{c_1}^{(22)} &=  \frac{3 \left(-9 g^4+6 g^2+1\right) \mn^3}{16 F^4}\ ,\quad \beta_{c_1}^{(21)}  =\frac{\left(-38 g^4+34 g^2+1\right) \mn^3}{32 F^4}\ ,\\
\beta_{e_m}^{(22)} &= -\frac{3 \left(17 g^4-26 g^2+3\right) \mn}{32 F^4}\ ,\quad \beta_{e_m}^{(21)}  =-\frac{27 \left(g^2-1\right)^2 \mn}{64 F^4}\ .
\end{align}

Furthermore, to remove the UV divergences from the one loop diagrams up to $\mathcal{O}(p^5)$, we get
\begin{align}
\beta_m^{(11)}&=\frac{3 g^2 \mn^3}{2 F^2}\ ,\\
\beta_{c_1}^{(11)}&=-\frac{3 g^2 \mn}{8 F^2}
+\frac{3 g^2 \ell_4 \mn^3}{4 F^4}-\frac{3 g \mn^3 (2 d_{16}-d_{18})}{2 F^2}\ , \\
\beta_{e_m}^{(11)}&=
\frac{3 (8 c_1-c_2-4 c_3)}{8 F^2}
+\frac{3 g^2 \mn (\ell_4-\ell_3)}{2 F^4}+\frac{3 g \mn (d_{18}-2 d_{16})}{F^2}\ .
\end{align}

Before ending this subsection, it is important to emphasize that the one-loop diagrams, acting as sub-diagrams of the two-loop diagrams, will also generate finite contributions of the type
\begin{align}\label{eq.sub.diag.con}
m_N^\text{sub-diag.}\sim \sum_{X}\bigg\{\bigg(-\frac{\beta_X^{(1)}}{ \epsilon\Lambda}\bigg) \times \big[\text{linear term in $\epsilon$ from one loops}\big]\bigg\}\ ,
\end{align}
with $\beta_{X}^{(1)}$ given in Eq.~\eqref{eq.EOMS.UV.beta}. These contributions will also partially cancel the two-loop PCB terms, which will be handled in the next subsection.

\subsection{PCB term subtraction\label{sec.PCB.sub}}

In baryon ChPT, a notable issue is caused by the non-zero nucleon mass in the chiral limit~\cite{Gasser:1987rb}. That is, when the nucleon appears as an internal propagator in loops, a direct application of the dimensional regularization with usual $\overline{\text{MS}}$ subtraction scheme leads to terms with chiral dimensions lower than the chiral order counted with the naive power-counting rule, for which a nucleon propagator is counted as $\mathcal{O}(p^{-1})$. Such terms are called the PCB terms. 
The EOMS scheme has been proven to be successful to restore the chiral power counting at the one-loop level, see e.g. Refs.~\cite{Bernard:2007zu, Scherer:2012xha,Geng:2013xn}. 

In the current two-loop case, we will perform the chiral expansion of the master integrals with the help of dimensional counting analysis method~\cite{Gegelia:1999qt,Smirnov:2002pj}, which is detailed in Appendices~\ref{sec.dim.counting.analysis} and~\ref{sec.exp.master.int}. However, the application of EOMS scheme at the two-loop level is quite challenging for two reasons. 
First, in addition to polynomial-type PCB terms, there are also PCB terms containing $\ln M$ from two loops, which originate from the region where one integration momentum is soft and the other is hard~\cite{Djukanovic:2015gna}. Physically, such terms are similar to the non-local UV divergences discussed above, thus will be called non-local PCB terms throughout this work.
Second, one must take into account the $\epsilon$ term of the LECs appearing in the one-loop diagrams, which will be compensated by the $1/\epsilon$ part from the loop integrals to give additional finite contributions. To obtain them, one can calculate, for instance, $\pi N$ scattering up to one loop by explicitly keeping the $\mathcal{\epsilon}$ terms. Based on the $\pi N$ scattering results in Ref.~\cite{Yao:2016vbz}, we establish the following decomposition of the relevant UV-renormalized LECs involved in the one-loop diagrams with chiral orders lower than $\mathcal{O}(p^5)$: 
\begin{align}\label{eq.olo_lec.shift}
X^r = \widetilde{X} + \frac{\widetilde{\beta}_X^{(0)}\mn}{\Lambda F^2} + \frac{\widetilde{\beta}_X^{(1)}\mn}{\Lambda F^2}\epsilon \ ,\quad X\in\{g,m,c_1,c_2,c_3\}\ ,
\end{align}
where the EOMS $\beta$ functions read
\begin{align}
\widetilde{\beta}_{g}^{(0)}=g^3 m \ , \quad \widetilde{\beta}_{m}^{(0)}=0 \ , \quad  \widetilde{\beta}_{c_1}^{(0)}=\frac{3}{8}g^2 \ ,\quad 
\widetilde{\beta}_{c_2}^{(0)}=-\frac{2+g^4}{2} \ , \quad \widetilde{\beta}_{c_3}^{(0)}=\frac{9g^4}{4} \ ,
\\
\widetilde{\beta}_{g}^{(1)}=0 \ , \quad \widetilde{\beta}_{m}^{(1)}=0 \ , \quad \widetilde{\beta}_{c_1}^{(1)}=0  \ , \quad 
\widetilde{\beta}_{c_2}^{(1)}=-\frac{54+\pi^2}{24} \ , \quad 
\widetilde{\beta}_{c_3}^{(1)}=-\frac{6+\pi^2}{48} \ .
\end{align}
Note that the LECs $\ell_{3,4}$ and $d_{16,18}$ in the $\mathcal{O}(p^5)$ one-loop diagrams remain untouched, since no PCB terms occur at this chiral order. We have explicitly verified that the two-loop non-local PCB terms are exactly canceled by counterparts from Eq.~\eqref{eq.sub.diag.con} and Eq.~\eqref{eq.olo_lec.shift}.

In fact, the above procedure can be conducted more compactly in $d$-dimensional spacetime. Namely, the LECs are split by means of
\begin{align}\label{eq.eoms.shift.d}
X^r = \widetilde{X} + \delta^{(1)}_X  \ ,\quad X\in\{g, m,c_1,c_2,c_3\} \ ,
\end{align}
where the $d$-dimensional shifts are taken from~\cite{Yao:2016vbz} 
\begin{align}
   \delta^{(1)}_{g}&= -\frac{g \left((d-2)g^2 - 4\right) }{2F^2}R^{\rm reg.}_{01}\ ,\\
   \delta^{(1)}_{m}&=-\frac{3 g^2 m }{2 F^2}R^{\rm reg.}_{01}\ ,\\
   \delta^{(1)}_{c_1}&=\frac{3 \, (d - 2) \, g^2 }{16 \, (d - 3) \, F^2 \, m}R^{\rm reg.}_{01} \ ,\\
   \delta^{(1)}_{c_2}&= -\frac{((d^2 - 6d + 12)g^4 - 8(d - 3)g^2 + 4)}{8(d-3)F^2m}R^{\rm reg.}_{01}\ ,\\
   \delta^{(1)}_{c_3}&= - \frac{1}{16 (d - 3) F^2 m} \bigg[3 d^3 g^4 - d^2 (21 g^2 + 8) g^2 \notag\\
   &\hspace{3cm}+ d (46 g^4 + 40 g^2 + 4) - 4 (9  g^4 + 14  g^2 + 3)\bigg]R^{\rm reg.}_{01} \ .
\end{align}
The analytical expression of $R^{\rm reg.}_{01}$ is presented in Eq.~\eqref{eq.ono.exp.reg} in Appendix~\ref{sec.exp.master.int}. By expanding the above shifts around $d=4$ up to $\mathcal{O}(\epsilon)$, one can reproduce the UV-$\beta$ functions and EOMS shifts for the LECs $X\in\{g, m,c_1,c_2,c_3\}$ in Eq.~\eqref{eq45} and Eq.~\eqref{eq.olo_lec.shift}, respectively. In $d$-dimensional spacetime, it is also easier to identify the non-local PCB terms, which are accompanied by a uniform factor of $(M/m)^{-2\epsilon}$. It is found that the shifts in Eq.~\eqref{eq.eoms.shift.d} exactly cancel the two-loop non-local PCB terms, which is again a nontrivial check of our results.

Once the non-local PCB terms from two loops are canceled by the one-loop subdiagrams, the remaining PCB terms are merely polynomials of the pion mass that can be absorbed in tree-level counter terms. Specifically, the UV-renormalized parameters in the tree-level counter terms are further decomposed as
\begin{align}
   X^r = \widetilde{X} + \delta^{(21)}_X +  \delta^{(22)}_X\ ,\quad X\in\{m, c_1, e_m\}\ ,
\end{align}
where the terms $\delta^{(21)}_X$ and $\delta^{(22)}_X$ ought to cancel the PCB terms stemming from the one- and two-loop diagrams, respectively. In $d$-dimensional space-time, the shifts $\delta^{(21)}_X$ read
\begin{align}
\delta_{m}^{(21)}&=-\frac{3g^2 m}{2F^2}R^{\rm reg.}_{01} \ , \\
\delta_{c_1}^{(21)}&=\frac{3g}{16F^4 m}\left\{
F^2\left[8(2d_{16}-d_{18})m^2+\frac{d-2}{d-3}g\right]
-4g \ell_4 m^2
\right\} R^{\rm reg.}_{01}\ ,
\\
\delta_{e_m}^{(21)}&=\frac{3(d-2)g}{4(d-3)F^4 m}\left[2F^2(2d_{16}-d_{18})+g(\ell_3-\ell_4)\right] R^{\rm reg.}_{01} \ .
\end{align}
For the $4$-dimensional spacetime, the above shifts are expanded around $d=4$ and only the finite pieces should be kept. This operation also applies to the shifts $\delta^{(22)}_X$ responsible for canceling two-loop PCB terms. The results of $\delta^{(22)}_X$ are
\begin{align}
\delta_{m}^{(22)}&=\frac{m}{2F^4}\bigg\{
3g^2[(d-2)g^2-1](R^{\rm reg.}_{01})^2
-m^2\big[3g^2 R^{\rm reg.}_{00111}
-\frac{2d-5}{(d-2)(3d-4)}
\notag\\
&\times [2(d-1)+4(2d-3)g^2+(5d-6)g^4] R^{\rm reg.}_{11100}
\big]
\bigg\} \ , \\
\delta_{c_1}^{(22)}&=-\frac{1}{16F^4 m}
\bigg\{
\frac{3(d-2)g^2}{(d-4)(d-3)}[2(d-4)+(d(d-3)g^2-1)](R^{\rm reg.}_{01})^2
\notag\\
&+m^2\left[ 
\frac{3(3d-8)g^4}{d-4}R^{\rm reg.}_{00111}
+[2(d-1)+4(2d-5)g^2+5(d-4)g^4]R^{\rm reg.}_{11100}
\right]
\bigg\} \ , \\
\delta_{e_m}^{(22)}&=-\frac{3}{128(d-4)^2F^4 m^3}
\bigg\{
\frac{(d-2)g^2}{(d-6)(d-5)(d-3)^2}
\bigg[ 
8(d-6)(d-4)^2(48+d(5d-33))
\notag\\
&
-(49176+d(d(43890+d(d(2078+d(4d-157))-13187))-73764))g^2 (R^{\rm reg.}_{01})^2
\bigg]
\notag\\
&-2m^2\bigg[ 
\frac{(3d-14)(3d-10)(3d-8)g^4}{d-6}R^{\rm reg.}_{00111}
-\frac{2(d-4)}{(d-3)(2d-7)}
\big[ 
2(d-4)(d-3)^3(d-1)
\notag\\
&
+4(d-4)(2d-7)(d(24+(d-9)d)-19)g^2\notag\\
&+(d-3)(952+d(d(479+d(5d-84))-1132))g^4 R^{\rm reg.}_{11100}
\big]
\bigg]
\bigg\}\ ,
\end{align}
where the two-loop regular integrals, $R^{\rm reg.}_{00111}$ and $R^{\rm reg.}_{11100}$ are given by Eq.~\eqref{eq.reg.00111} and~\eqref{eq.reg.11100} in Appendix~\ref{sec.exp.master.int}, respectively.

\subsection{Full EOMS-renormalized nucleon mass\label{sec.full.EOMS.mn}}
Eventually, the full EOMS-renormalized nucleon mass up to $\mathcal{O}(p^5)$ can be summarized as
\begin{align}\label{eq.full.form.mn}
    m_N = \widetilde{m} -\underbrace{4 \widetilde{c}_1 M^2}_{\mathcal{O}(p^2)} + \underbrace{\bar{m}_N^{(1a)}}_{\mathcal{O}(p^3)} -\underbrace{2 \widetilde{e}_m M^4 +\bar{m}_N^{(1b)}}_{\mathcal{O}(p^4)} + \underbrace{\bar{m}_N^{(1c)}+ 2 \bar{m}_N^{(1d)} + \bar{m}_N^\text{2-loop}+\bar{m}_N^\text{sub-diag.}}_{\mathcal{O}(p^5)} \ .
\end{align}
A bar over the loop contribution means that the UV divergences and PCB terms are subtracted. The quantities $\widetilde{m}$, $\widetilde{c}_1$ and $\widetilde{e}_m$ stand for the two-loop EOMS-renormalized parameters. The term $\bar{m}_N^\text{2-loop}$ is the sum of all the subtracted two-loop contributions corresponding to the diagrams in Fig.~\ref{fig:2loop} and the piece of Eq.~\eqref{eq:1loto2lo}. The last term of Eq.~\eqref{eq.full.form.mn} is nothing but the power counting preserving part of Eq.~\eqref{eq.sub.diag.con}. 

\begin{figure}[ht]
    \centering
   \includegraphics[width=0.99\textwidth]{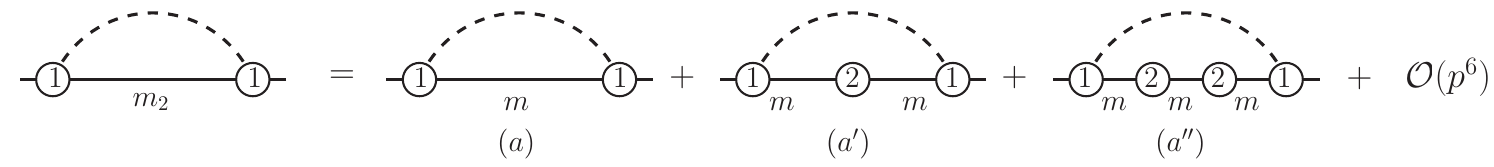}
    \caption{Illustration of generating one-loop diagrams with mass insertions.}
    \label{fig:mn.mass.insertion}
\end{figure}

It should be emphasized that, in Eq.~\eqref{eq.full.form.mn}, the nucleon mass parameter in the loop corrections is $\widetilde{m}_4=\widetilde{m}-4\widetilde{c}_1 M^2-2\widetilde{e}_m M^4$. 
From the viewpoint of perturbation theory, it is more customary to re-express $\widetilde{m}_4$ by $\widetilde{m}$. In fact, in the loop contributions the two parameters are interchangeable.
Firstly, the $\widetilde{m}_4$ in the $\mathcal{O}(p^5)$ loops may be replaced directly by $\widetilde{m}$, since the difference caused is of higher order beyond our working accuracy. Secondly, the final expression of $\bar{m}_N^{(1b)}$ at $\mathcal{O}(p^4)$ is independent of $\widetilde{m}_4$. Finally, for $\bar{m}_N^{(1a)}$ at $\mathcal{O}(p^3)$, let us write down the explicit $m_4$-dependence as,
\begin{align}
    {m}_N^{(1a)} =\Sigma_N^{(1a)}({m}_4,{m}_4)\ .
\end{align}
The substitution of ${m}_4$ by ${m}$ can be done in two steps. The second argument $m_4$ is responsible for the nucleon mass parameter in the propagator; therefore, changing this $m_4$ to $m$ generates diagrams of $\mathcal{O}(p^4)$ and $\mathcal{O}(p^5)$, as illustrated in Fig.~\ref{fig:mn.mass.insertion}, which are referred to as mass-insertion contributions. On the other hand, one should further replace the first argument $m_4$ by $m$, which leads to the so-called off-shell effect, similar to Eq.~\eqref{eq:1loto2lo}. Fortunately, it can be proved that the two kinds of induced contributions cancel each other out.

More specifically, let us perform Taylor expansion of $\bar{m}_N^{(1a)}$ around $m$, 
\begin{align}
    \Sigma_N^{(1a)}(m_4,m_4) &= \Sigma_N^{(1a)}(m,m) + (m_2-m)\big[\Sigma^{[1,0]}(m,m)+ \Sigma^{[0,1]}(m,m)\big] \notag\\
    & +\frac{1}{2}(m_2-m)^2\big[\Sigma^{[2,0]}(m,m)+2\Sigma^{[1,1]}(m,m)+\Sigma^{[0,2]}(m,m)]+\mathcal{O}(p^6)\ ,\label{eq.se.m2.m2}
\end{align}
where $\Sigma^{[i,j]}$ denotes derivatives as~\footnote{Here we use $\texttt{m}$ to denote a general mass parameter.} 
\begin{align}
   \Sigma^{[i,j]}(\slashed{p},\texttt{m}) =\left(\frac{\partial}{\partial \slashed{p}}\right)^i \left(\frac{\partial}{\partial \texttt{m}}\right)^j \Sigma^{(1a)}_N(\slashed{p},\texttt{m}) \ ,
\end{align}
with $\Sigma^{(1a)}_N(\slashed{p},\texttt{m})$ the self-energy corresponding to diagram $(a)$ of Fig.~\ref{fig:1loop}. We write
\begin{align}
 \Sigma^{(1a)}_N(\slashed{p},\texttt{m})=   \frac{3g^2\kappa}{4F^2}\int\frac{{\rm d}^d \ell}{(i\pi^{d/2})}
\bigg\{\left(\slashed{\ell}\gamma_5\right)
\frac{1}{[\ell^2-M^2][(\slashed{\ell}+\slashed{p})-\texttt{m}]}\left(\slashed{\ell}\gamma_5\right)
\bigg\} \ ,
\end{align}
and it can be seen that only the nucleon propagator depends on $\slashed{p}$ and $\texttt{m}$. The derivative acting on the nucleon propagator reads
\begin{align}
    \left(\frac{\partial}{\partial \slashed{p}}\right)^i \left(\frac{\partial}{\partial \texttt{m}}\right)^j\frac{1}{(\slashed{\ell}+\slashed{p})-\texttt{m}}=\frac{(-1)^j}{[(\slashed{\ell}+\slashed{p})-\texttt{m}]^{1+i+j}}\ .
\end{align}
Therefore, one readily finds
\begin{align}
    &\Sigma^{[1,0]}(m,m) = -\Sigma^{[0,1]}(m,m)\ ,\\
    &\Sigma^{[2,0]}(m,m)  =\Sigma^{[0,2]}(m,m)  = -\Sigma^{[1,1]}(m,m) \ .
\end{align}
Taking the above two identities  back to Eq.~\eqref{eq.se.m2.m2}, we obtain
\begin{align}
   \Sigma_N^{(1a)}(m_4,m_4) = \Sigma_N^{(1a)}(m,m) \ ,
\end{align}
up to higher order contribution beyond $\mathcal{O}(p^5)$. That is, $m_4$ in $m_N^{(1a)}$ can be replaced by $m$ directly. In fact, such a cancellation between the mass-insertion and off-shell self-energy contributions has been demonstrated explicitly in the heavy baryon ChPT calculation as well in Ref.~\cite{McGovern:1998tm}, see Eqs.~(7) and~(8) therein.

In summary, the nucleon mass $\widetilde{m}_4$ in the full EOMS result presented in Eq.~\eqref{eq.full.form.mn} can be treated as the one in the chiral limit, i.e. $\widetilde{m}$. Hereafter, we will suppress the widetildes over the renormalized parameters for simplicity if no confusion is caused.

\subsection{Chiral Expansion truncated at $\mathcal{O}(p^5)$\label{sec.chiral.exp}}

The chiral expansion of the nucleon mass up to $\mathcal{O}(p^5)$ can be written in the form as
\begin{align}\label{eq.ce.p5}
 m_N = m+ k_1 M^2 + k_2 M^3 + k_3 M^4\ln\frac{M}{\mu}
 +k_4 M^4 + k_5 M^5\ln\frac{M}{\mu}
 +k_6 M^5 \ .
\end{align}
We can derive analytical expressions for coefficients $k_i$'s with the help of the formulae presented in Appendices~\ref{sec.dim.counting.analysis} and~\ref{sec.exp.master.int}. Within various renormalization schemes, some of the $k_i$ coefficients are usually not the same, as can be seen below. In the EOMS scheme, these coefficients are given by
\begin{align}
    k_1^{\rm EOMS} &= -4 c_1 \ ,\notag\\
    k_2^{\rm EOMS} &= -\frac{3g^2\pi}{ 2 F^2 \Lambda}\ ,\notag\\
    k_3^{\rm EOMS} & =-\frac{3  {g}^2}{2  F^2 {m}\Lambda}+\frac{3   (8 {c_1}-{c_2}-4 {c_3})}{2  F^2\Lambda} \ ,\notag\\
    k_4^{\rm EOMS} & = -2 e_m +\frac{3 g^2 }{2  F^2 m\Lambda}+\frac{3 c_2 }{8  F^2\Lambda} \ ,\notag\\
    k_5^{\rm EOMS} & = \frac{3 {g}^2 (16 g^2-3)\pi }{4  F^4 \Lambda^2}\ ,\notag\\
    k_6^{\rm EOMS} & =\frac{19 \pi  g^4}{4 F^4 \Lambda^2}+\frac{6 \pi  g (d_{18}-2 d_{16})}{F^2 \Lambda}+{g}^2 \left[\frac{6 \pi }{F^4 \Lambda^2}+\frac{3 \pi  \left(F^2+8 m^2 (2 \ell_4-3 \ell_3)\right)}{16 F^4 \Lambda m^2}\right]\ . \label{eq.ki.EOMS}
\end{align}
The chiral expansion unavoidably enforces a truncation at a certain order, which changes the analytic property (as a function of $M$) 
of the original expression. For instance, one expects the emergence of a di-logarithm term, $[\ln(M/\mu)]^2$, in a two-loop calculation. The di-logarithm term does exist in the original expression. However, such term is absent in Eq.~\eqref{eq.ce.p5} as it is at least of $\mathcal{O}(p^6)$. 

In the IR prescription~\cite{Ellis:1997kc, Becher:1999he}, the regular parts of the loop integrals are dropped to restore the power counting. Specifically, in our case, the one- and two-loop integrals, $J_{\nu_1\nu_2}$ and $I_{\nu_1\nu_2\nu_3\nu_4\nu_5}$, are set to their purely infrared singular parts, $J_{\nu_1\nu_2}^{(1)}$ and $I_{\nu_1\nu_2\nu_3\nu_4\nu_5}^{(1,1)}$ (cf. Eqs.~\eqref{eq.ce.olo.master.int} and~\eqref{eq.ce.tlo.master.int}). Or equivalently and conveniently, one just sets all the encountered regular master integrals, as shown in Appendix~\ref{sec.exp.master.int}, to zero. 
However, a subtlety that should be taken with special caution is the contribution of the term $m_N^\text{sub-diag.}$ in Eq.~\eqref{eq.full.form.mn}. One should utilize the UV-beta functions obtained in IR prescription for the LECs in the one-loop diagrams, which are~\cite{Siemens:2016hdi}
\begin{align}
    \beta_m^{(1)}=\beta_g^{(1)}=\beta_{c_{1,2,3}}^{(1)}
    =\beta_{d_{18}}^{(1)}=0\ ,\,\beta_{\ell_3}^{(1)}=- \frac{1}{4}\ , \,\beta_{\ell_4}^{(1)}=1\ ,\,
 \beta_{d_{16}}^{(1)} = \frac{g(1 + 2g^2)}{4 F^2 }\ ,
\end{align}
instead of the ones in Eq.~\eqref{eq.EOMS.UV.beta}. Eventually, the chirally expanded nucleon mass in IR prescription up to $\mathcal{O}(p^5)$ has the same form as Eq.~\eqref{eq.ce.p5} but with the following $k_i$ factors:
\begin{align}
    k_1^{\rm IR} &= -4 c_1 \ ,\notag\\
    k_2^{\rm IR} &= -\frac{3g^2\pi}{2 F^2\Lambda}\ ,\notag\\
    k_3^{\rm IR} & =-\frac{3  {g}^2}{2 F^2 {m}\Lambda}+\frac{3   (8 {c_1}-{c_2}-4 {c_3})}{2 F^2\Lambda} \ ,\notag\\
    k_4^{\rm IR} & = -2 e_m -\frac{3 g^2 }{4 F^2 m\Lambda}+\frac{3 c_2 }{8  F^2\Lambda} \ ,\notag\\
    k_5^{\rm IR} & = \frac{3 {g}^2 (16 g^2-3)\pi }{4  F^4 \Lambda^2}\ , \notag\\
    k_6^{\rm IR} & =\frac{3 \pi  g^4}{F^4 \Lambda^2}
    +\frac{6 \pi  g (d_{18}-2 d_{16})}{F^2 \Lambda}+\frac{3 \pi  g^2 \left(F^2+8 m^2 (2 \ell_4-3 \ell_3)\right)}{16 F^4 \Lambda m^2}\ .    \label{eq.ki.IR}
\end{align}

We have reproduced the coefficients obtained previously in Refs.~\cite{Schindler:2006ha,Schindler:2007dr}. Note that, the expanded IR results are actually identical to the HB ones, given that the $1/m$ corrections are included (cf. Refs.~\cite{Bernard:1995dp,Fettes:2000gb}). Specifically, the one-loop $k_{1-4}^{\rm IR}$ and two-loop $k_{5,6}^{\rm IR}$ coefficients coincide with those derived in HBChPT calculations in Refs.~\cite{Fettes:2000xg} and~\cite{McGovern:1998tm}, respectively. 
Interestingly, the leading non-analytic terms in quark masses, associated with $k_2$, $k_3$ and $k_5$, are the same in EOMS and IR (or HB). 
However, the $k_6$ term, which is also non-analytic, depends on renormalization schemes. This is due to the fact that the one-loop sub-diagrams can contribute to $k_6$, and different treatments of the LECs in the one-loop sub-diagrams lead to different $k_6$. In Ref.~\cite{Bijnens:2014ila}, the method of renormalization group equation is employed to derive the leading logarithms of the nucleon mass by using one-loop beta functions as inputs. 
The renormalization scale invariance  guarantees that the leading chiral logarithm terms at each chiral order, such as $k_3$ and $k_5$, must be scheme independent.\footnote{In Ref.~\cite{Bijnens:2014lea}, the leading logarithm expansion of the nucleon mass is given by $m_N=m+\frac{M^2}{m}\sum_{n=1}^\infty k_{2n}L^{n-1}+\pi M\frac{M^2}{(4\pi F)^2}\sum_{n=1}^\infty k_{2n+1}L^{n-1}$ with the chiral logarithm $L=\frac{M^2}{(4\pi F^2)}\ln\frac{\mu^2}{M^2}$ and the coefficients $k_i$ compiled in Table~1 of  Ref.~\cite{Bijnens:2014lea}. It is verified that different parametrizations of the effective Langrangians yield the same result. Various renormalization schemes actually correspond to redefinitions of fields in Lagrangians. That is, the leading logarithms are independent of renormalization schemes.}

\section{Numerical results and discussions\label{sec.numerical.result}}
The full EOMS expression of the nucleon mass in Eq.~\eqref{eq.full.form.mn} possesses original analytical properties from the covariant quantum field theory and hence is appropriate and reliable to perform chiral extrapolation to unphysical pion masses. In this section, we confront our chiral result with lattice QCD data as the pion mass changes.

\subsection{Parameter setup}
In our numerical calculation, the two-loop master integrals are computed using the package \texttt{AMFlow}~\cite{Liu:2022mfb,Liu:2022tji,Liu:2022chg} and the renormalization scale is set to $\mu=m_N$. Note that in the limit of the exact isospin symmetry of SU(2), we always choose the physical mass of the nucleon to be $ m_N = 939$~MeV and the physical mass of the pion to be $M_\pi=138$~MeV. These values are the averaged masses of the physical states in the nucleon iso-doublet and the pion iso-triplet, respectively. 
The parameters $g$ and $F$ from the leading-order nucleonic and purely mesonic Lagrangians are axial couplings in the chiral limit. Therefore, they are quark-mass independent and can be set as constants: $g = 1.13$~\cite{Yao:2017fym} and $F=$~86.7~MeV~\cite{FlavourLatticeAveragingGroupFLAG:2021npn}\footnote{The recent lattce QCD results of the nucleon mass~\cite{Hu:2024mas}, to which we will perform fits, are obtained with $N_f=2+1$ flavor QCD ensembles. Accordingly, we prefer to use the $N_f=2+1$ FLAG average of $F_\pi/F=1.062(2)$ summaried in Ref.~\cite{FlavourLatticeAveragingGroupFLAG:2021npn} to determine the parameter $F$.}. Equivalently, one may replace the two parameters by their physical counterparts, namely $g_A=1.267$ and $F_\pi=92.1$~MeV~\cite{ParticleDataGroup:2024cfk}, e.g., through the one-loop relations of Eqs.~\eqref{eq.gA.1loop} and \eqref{eq.Fpi.1loop}, respectively. However, such a substitution does not bring any convenience because one has to consider the pion-mass dependence of the physical $g_A$ and $F_\pi$ additionally. Therefore, we stay with using the chiral limit quantities $g$ and $F$. The other one-loop renormalized LECs occurring in higher-order $\pi N$ Lagrangians are set as follows~\cite{Siemens:2016jwj,Yao:2017fym,Chen:2012nx}:
\begin{align}
   & c_2= 3.35\pm 0.03~\textrm{GeV}^{-1}\ ,\quad c_3=-5.70\pm 0.06~\textrm{GeV}^{-1}\ ,\notag\\ 
    &d_{16}= -0.83
    \pm 0.03~\textrm{GeV}^{-2}\ ,\quad 
    d_{18} =-0.56\pm 1.42~\textrm{GeV}^{-2}\ ,
\end{align}
which are independent of the pion masses. 
Here $c_{2,3}$ and $d_{18}$ are determined by fitting to $\pi N$ scattering phase shifts in Ref.~\cite{Chen:2012nx,Siemens:2016jwj}, while $d_{16}$ is obtained by confronting the chiral prediction of axial form factors with lattice QCD data~\cite{Yao:2017fym}. The $\mathcal{O}(p^4)$ mesonic LECs $\ell_3$ and $\ell_4$ can be found in the classical work by Gasser and Leutwyler~\cite{Gasser:1983yg}, and updates on their values are summarized in the comprehensive reviews~\cite{Bijnens:2014lea,FlavourLatticeAveragingGroupFLAG:2021npn}. The values of $\ell_{3,4}$ at $\mu=m_N$ can be obtained by utilizing the evolution equations~\cite{Gasser:1983yg}
\begin{align}
    \ell_i^r(\mu) =\frac{\beta^{(1)}_{\ell_i}}{16\pi^2}\bigg[\bar{\ell}_i+\ln\frac{M_\pi^2}{\mu^2}\bigg] ,\quad i=3,4
\end{align}
with $\beta^{(1)}_{\ell_i}$ given in Eq.~\eqref{eq.EOMS.UV.beta}. Taking  the recent FLAG averages ($N_f=2+1$)~\cite{FlavourLatticeAveragingGroupFLAG:2021npn}, $\bar{\ell}_3=3.07(64)$ and $\bar{\ell}_4=4.02(45)$, as inputs, one obtains 
\begin{align}
   10^{3}\,\ell_3^r(\mu=m_N)= 1.21\pm 1.01\ ,\quad
   10^{3}\,\ell_4^r(\mu=m_N)= 1.17\pm 2.85\ .
\end{align} 
The remaining parameters are $m$, $c_1$, and $e_m$, which are renormalized at the two-loop level and will be taken as free parameters in the fit.

\subsection{Description of lattice QCD data}
The nucleon mass has been extensively computed on the lattice by various groups, e.g., see Refs.~\cite{BMW:2008jgk,Alexandrou:2014sha,Lin:2019pia,Hu:2024mas}. Here we focus on a test of the convergence property of the two-loop chiral result of the nucleon mass by analyzing the most recent lattice QCD data~\cite{Hu:2024mas} with varying pion masses. 

\begin{figure}[t]
\centering
\includegraphics[width=0.7\textwidth]{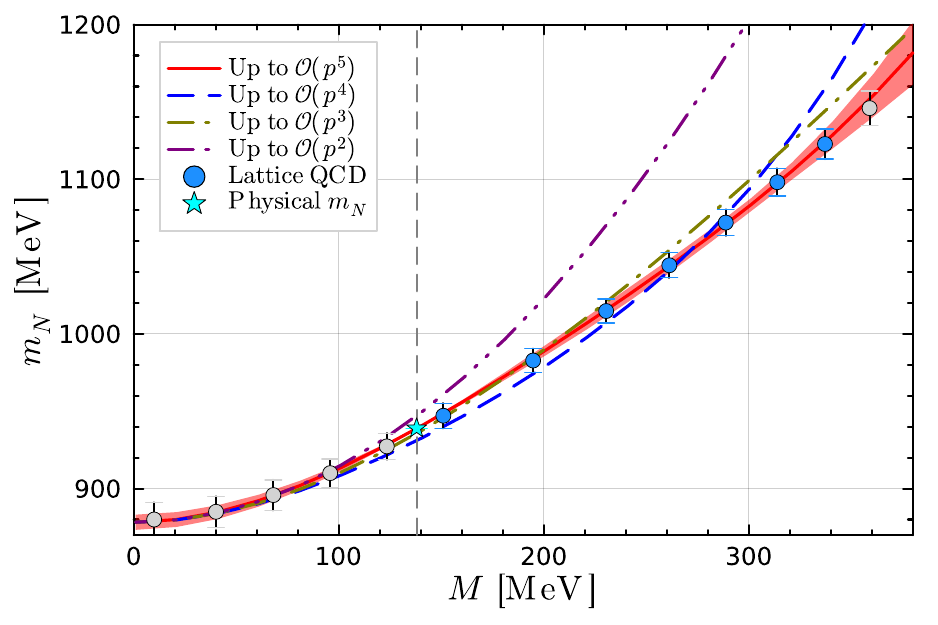}
\caption{Pion mass dependence of the nucleon mass. Our leading two-loop result (up to $\mathcal{O}(p^5)$) is represented by the red solid line. The recent lattice QCD results, taken from the continuum extrapolation in Ref.~\cite{Hu:2024mas}, are denoted by the blue and gray filled circles, where only the blue ones are fitted. The cyan star stands for the value of the nucleon mass at physical pion mass, i.e. $m_N^\text{PDG}$~\cite{ParticleDataGroup:2024cfk}. The chiral results up to $\mathcal{O}(p^4)$, $\mathcal{O}(p^3)$, and $\mathcal{O}(p^2)$, computed using the same LECs as the two-loop result, are shown for easy comparison as well. The dashed gray vertical line shows the physical value of the pion mass, $M_\pi = 138$~MeV.}
\label{fig:chiral.exp.mn}
\end{figure}

There are three free two-loop renormalized parameters in our full EOMS expression: $m$, $c_1$, and $e_m$. 
Here, we choose to pin down their values by performing a simultaneous fit to the very recent lattice QCD results by the CLQCD Collaboration~\cite{Hu:2024mas} and the physical value of the nucleon mass~\cite{ParticleDataGroup:2024cfk}. Seven points in the pion mass range $\sim[150,340]$~MeV taken from the continuum extrapolation in Ref.~\cite{Hu:2024mas} are fitted, which are shown as blue circles in Fig.~\ref{fig:chiral.exp.mn}.
Since the ChPT calculation in this work is performed in the limit of exact isospin symmetry, we use $m_N^{\rm PDG} = (938.92\pm 0.65) $~MeV for the experimental nucleon mass. Here, the central value is set to the average mass of the proton and neutron, while the uncertainty is set to one-half of their mass difference. The fit results of the two-loop renormalized LECs are compiled in the second column of Table~\ref{tab:fit.LECs}. The $\chi^2$ per degrees of freedom is too small because the correlations of the fitted lattice QCD results are unknown to us.\footnote{The fits are carried out by setting $c_{2,3}$, $d_{16,18}$ and $\ell_{3,4}$ to their respective central values given in the above subsection. We have checked that stable fit results can be achieved by varying those parameters within their $1\sigma$ uncertainties. } 
Thus, a more robust error analysis can only be performed in the future when correlations among the lattice data points are available.
For comparison, 
results of Ref.~\cite{Chen:2024twu} are shown in the last column. 
\begin{table}[ht]
\renewcommand{\arraystretch}{1.3}
\caption{Fit results of the two-loop renormalized LECs. Results of Ref.~\cite{Chen:2024twu} are also shown for comparison and the asterisk denotes an input parameter.
}\label{tab:fit.LECs}
\begin{ruledtabular}
\begin{tabular}{l|ccc|c }
LECs & EOMS & EOMS-truncated & IR-truncated    & Ref.~\cite{Chen:2024twu} \\
\hline
$m$~ [MeV]  & $878.2\pm 5.0$ & $866.0\pm 5.0$ &$873.2\pm 5.0$  & $856.6\pm 1.7$\\
$c_1$~ [GeV$^{-1}$] & $-0.90\pm0.07$ & $-1.28\pm0.07$ & $-1.13\pm 0.07$  & $-1.39^\ast$\\
$e_m$ [GeV$^{-3}$] & $-7.41\pm 1.71$ & $7.75\pm1.72$ & $-3.61\pm 1.71$ & $3.42\pm 0.25$\\
\hline
${\chi^2}/{\rm d.o.f.}$& ${0.13}/{(7+1-3)}$ & ${1.12}/{(7+1-3)}$ & ${0.49}/{(7+1-3)}$ & $-$
\end{tabular}
\end{ruledtabular}
\end{table}

It can be seen that our two-loop determination yields $m=(878.2\pm5.0)$~MeV, which agrees well with the one-loop outcome $m=(882.8\pm4.0)$~MeV estimated by Ref.~\cite{Fuchs:2003kq}. This agreement is quite reasonable, as explained below. The derived chiral corrections are expressed in terms of the pion mass, and therefore their contribution to the nucleon mass must be zero in the chiral limit.
Or equivalently, the chiral limit mass $m$ is a LEC independent of the pion mass.
By matching the result to the physical value of the nucleon mass, a higher-order calculation of the nucleon mass only leads to a rearrangement of the contributions of various chiral orders in $\Delta m_N \equiv m_N-m$. As long as the chiral series converges well, the determined chiral limit mass $m$ should vary little with the chiral order.
The chiral limit nucleon mass $m$ mainly originates from the gluonic part of the trace anomaly in the hadron mass decomposition~\cite{Collins:1976yq,Nielsen:1977sy,Shifman:1978zn}, which was determined to be $\sim 880$~MeV by the very recent lattice QCD simulation conducted with seven pion masses $\in\{135,350\}$~MeV~\cite{Hu:2024mas}.
The lattice determination agrees well with our result. A smaller value $m=(856.6\pm1.7)$~MeV was obtained in Ref.~\cite{Chen:2024twu}, which should be mainly due to the use of the chirally truncated nucleon mass. By performing a fit with Eq.~\eqref{eq.ce.p5} and Eq.~\eqref{eq.ki.EOMS}, we get $m=(866.0\pm 5.0)$~MeV (shown in the third column of Table~\ref{tab:fit.LECs}), indicating that the truncation indeed leads to a downward shift of chiral-limit nucleon mass.

Here we determine the values of $c_1$ and $e_m$ at two-loop level with full EOMS chiral results for the first time. Their values turn out to be of natural size, satisfying the naturalness requirement of Wilson coefficients involved in effective field theories~\cite{Meissner:2022cbi}. 
The result of $c_1$ now accounts for the impact of two-loop corrections and decreases in size compared to the corresponding one-loop determinations given in e.g. Refs.~\cite{Chen:2012nx,Siemens:2016jwj}. 
The value of $e_m$ differs in sign from the one in Ref.~\cite{Chen:2024twu}, $e_m^{\rm Chen}\simeq 3.42\pm 0.25$,\footnote{ This value is obtained by using $\hat{e}_1 M^4=(2.7\pm 0.2)$~MeV in Ref.~\cite{Chen:2024twu} with $M=141$~MeV, where $\hat{e}_1=2e_m$.}
which was obtained by making use of an expanded chiral series truncated at $\mathcal{O}(p^6)$. We ascribe such a discrepancy to the change of the analytical structure of the original chiral result, as discussed in the text below Eq.~\eqref{eq.ki.EOMS}. To assess the difference between the full and expanded EOMS results, we have performed a comparative fit by utilizing the expanded chiral expression given in Eqs.~\eqref{eq.ce.p5} and~\eqref{eq.ki.EOMS}. 
The values of the fit parameters are shown in the third column of Table~\ref{tab:fit.LECs}. In this case, a positive $e_m$ is obtained. For comparison, the expanded IR result of Eqs.~\eqref{eq.ce.p5} and~\eqref{eq.ki.IR} is also used to fit the lattice QCD data points with the fitting results collected in the fourth column of Table~\ref{tab:fit.LECs}. A negative value of $e_m$ is observed within the IR prescription.  
More interestingly, the parameter $e_m$ is sensitive to the pion-nucleon sigma term. It is shown in Ref.~\cite{Scherer:2012xha} that the $\sigma_{\pi N}=(45\pm 8)$~MeV from the dispersive analysis in Ref.~\cite{Gasser:1990ce} leads to $-2\hat{e}_1M^4\approx (4.5\pm 8)$~MeV with $\hat{e}_1=2e_m$, implying a negative $e_m\approx (-3.1\pm 5.5)$~GeV$^{-3}$ albeit with large error. By the same token, the modern dispersive determination with a larger $\sigma_{\pi N}=59.1\pm 3.5$~\cite{Hoferichter:2015dsa} requires a more negative $e_m\approx (-12.8\pm 2.4)$~MeV.

\begin{figure}[t]
\centering 
\includegraphics[width=0.7\textwidth]{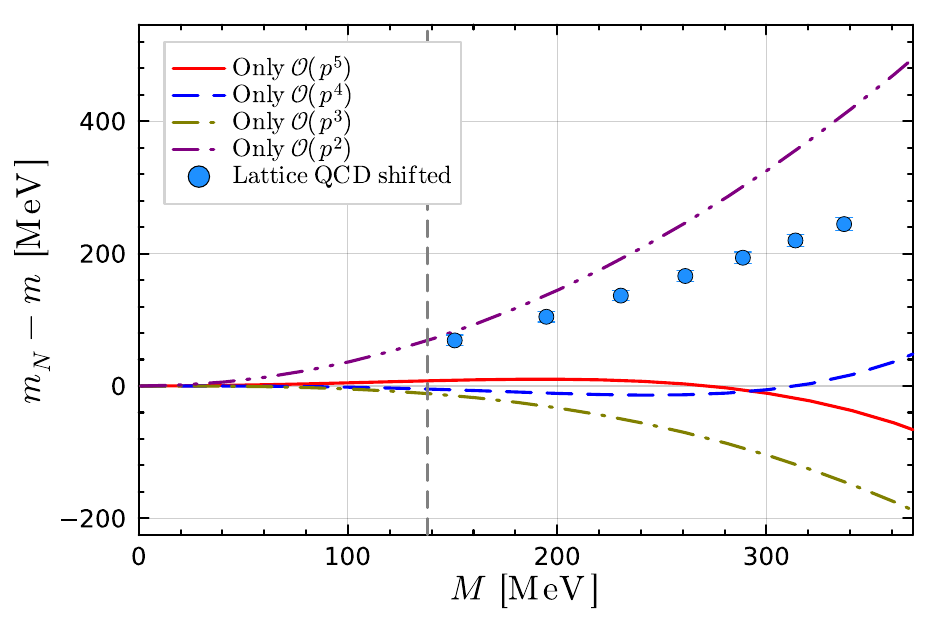}
\caption{Convergence property of the nucleon mass. Chiral corrections of different orders are shown separately. The lattice QCD data~\cite{Hu:2024mas} have been shifted downwards by $m$ to get the chiral corrections $m_N-m$.  }
\label{fig:chiral.exp.mn.orders}
\end{figure}

Based on the full-EOMS fit results, pion mass dependence of the nucleon mass is shown in Fig.~\ref{fig:chiral.exp.mn}. The curve of the $\mathcal{O}(p^5)$ nucleon mass is in excellent agreement with the lattice QCD data~\cite{Hu:2024mas} at various pion masses, even beyond the fit range $M\in[150,340]$~MeV. The band is obtained by varying the fit parameters within their $1\sigma$ uncertainties. In the same figure, we also show the pion mass dependence of the nucleon mass from the chiral expressions up to $\mathcal{O}(p^2)$, $\mathcal{O}(p^3)$ and $\mathcal{O}(p^4)$, with the same LECs as the two-loop result. The $\mathcal{O}(p^3)$ and $\mathcal{O}(p^4)$ results start to sizably deviate from the leading two-loop result at approximately $M_\pi=200$~MeV and $M_\pi=300$~MeV, respectively. 
The comparison in Fig.~\ref{fig:chiral.exp.mn} indicates a good convergence property of the EOMS result. 

To assess in more detail the convergence property of the EOMS results at different chiral orders, we plot the corrections of different orders in Fig.~\ref{fig:chiral.exp.mn.orders}. It is clearly observed that the EOMS chiral series of the nucleon mass converges fast below $M=300$~MeV. In the vicinity of $M=200$~MeV, there is a cancellation between the individual $\mathcal{O}(p^4)$ and $\mathcal{O}(p^5)$ contributions. 
Note that the contribution of $\mathcal{O}(p^5)$ to the nucleon mass is less than $10$~MeV for the pion mass $<300$~MeV. 
Therefore, in that region, the use of the complete one-loop result, i.e., up to $\mathcal{O}(p^4)$, for chiral extrapolation~\cite{Alvarez-Ruso:2013fza} is a good approximation. Beyond $M=300$~MeV, the two-loop contribution becomes sizable and indispensable for a proper and reliable chiral extrapolation. At the physical point, contributions from various chiral orders to the nucleon mass, corresponding to Eq.~\eqref{eq.full.form.mn}, read 
\begin{align}
    m_N =\big\{ 878.2 + \underbrace{68.8}_{\mathcal{O}(p^2)} + \underbrace{[- 11.4]}_{\mathcal{O}(p^3)} + \underbrace{[-4.6]}_{\mathcal{O}(p^4)} + \underbrace{7.9}_{\mathcal{O}(p^5)}\big\}~\textrm{MeV} = 938.9~\textrm{MeV}\ .
\end{align}

\begin{figure}[t]
\centering
\includegraphics[width=0.7\textwidth]{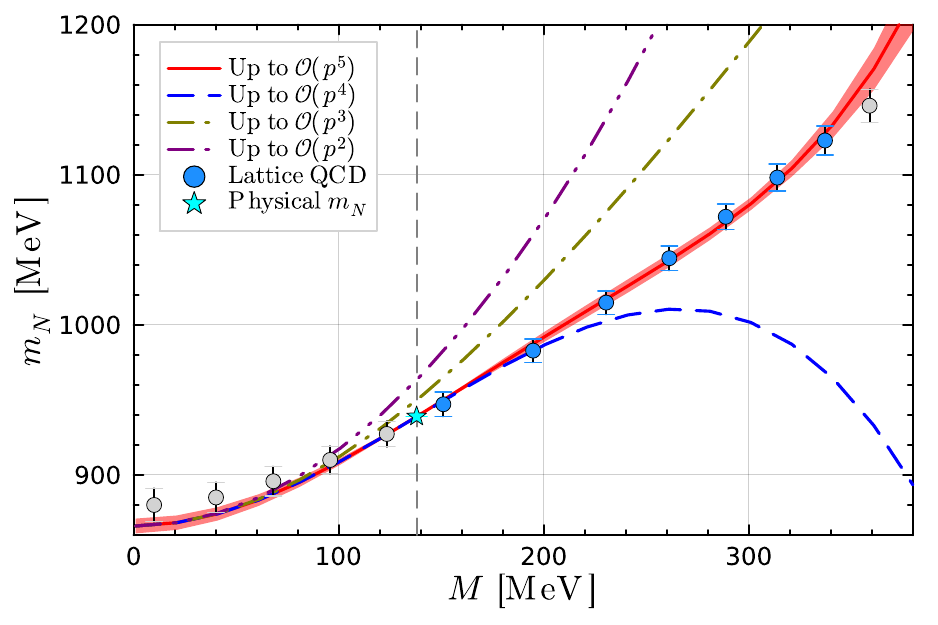}
\caption{Pion mass dependence of the nucleon mass within truncated EOMS approach. The leading two-loop result (up to $\mathcal{O}(p^5)$) is represented by the red solid line. The recent lattice QCD data, taken from Ref.~\cite{Hu:2024mas}, are denoted by the dodger-blue and gray filled circles, where only the dodger-blue ones are fitted. The cyan star stands for the value of the nucleon mass at physical pion mass, i.e. $m_N^\text{PDG}$~\cite{ParticleDataGroup:2024cfk}. The chiral results up to $\mathcal{O}(p^4)$, $\mathcal{O}(p^3)$, and $\mathcal{O}(p^2)$, computed using the same LECs as the two-loop result, are shown for easy comparison as well. The dashed gray vertical line shows the physical value of the pion mass, $M_\pi = 138$~MeV.}
\label{fig:chiral.exp.mn.eEOMS}
\end{figure}

\begin{figure}[t]
\centering 
\includegraphics[width=0.7\textwidth]{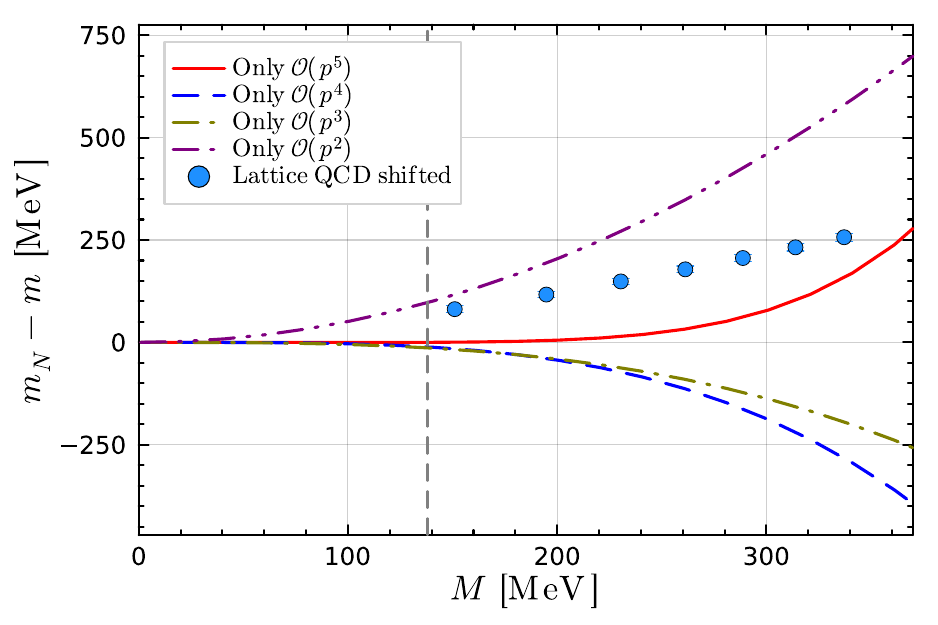}
\caption{Convergence property of the nucleon mass within truncated EOMS approach. Chiral corrections of different orders are shown separately. The lattice QCD data~\cite{Hu:2024mas} have been shifted downwards by $m$ to get the chiral corrections $m_N-m$. }
\label{fig:chiral.exp.mn.orders.eEOMS}
\end{figure}

For the truncated EOMS case, the behaviour of the pion-mass dependence is worse than the full EOMS one. One can clearly see from Figure~\ref{fig:chiral.exp.mn.eEOMS} that the truncated EOMS prediction starts to fail in describing the lattice QCD data beyond the fitting range, i.e. $M<70$~MeV and $M>340$~MeV. Furthermore, the convergency property of the chiral result becomes bad for unphysical pion mass $M>200$~MeV, as illustrated in Figure~\ref{fig:chiral.exp.mn.orders.eEOMS}. One may conclude that the expanded-EOMS chiral result is not suitable for a proper chiral extrapolation.

\begin{figure}[t]
\centering
\includegraphics[width=0.7\textwidth]{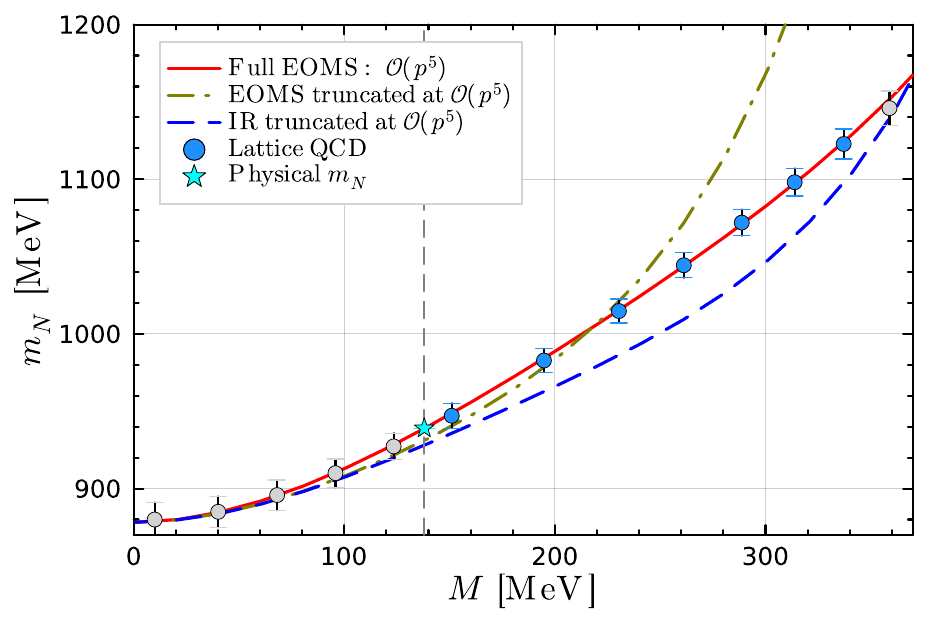}
\caption{Comparison of the nucleon mass obtained in full EOMS, EOMS truncated at $\mathcal{O}(p^5)$, and IR truncated at $\mathcal{O}(p^5)$. The Lattice QCD data, taken from Ref.~\cite{Hu:2024mas} are shown by dodger-blue or gray filled circles. The physical nucleon mass~\cite{ParticleDataGroup:2024cfk} is indicated by the cyan star.}
\label{fig:chiral.exp.mn.compare}
\end{figure}

It is instructive to compare the full-EOMS result with the truncated EOMS and IR ones, specified in Eq.~\eqref{eq.ki.EOMS} and~\eqref{eq.ki.IR}, respectively, by making use of a common set of LEC values as input. 
Figure~\ref{fig:chiral.exp.mn.compare} shows a comparison with varying the pion mass, based on the values of the LECs given in the second column of Table~\ref{tab:fit.LECs}. As mentioned above, the truncation of the EOMS result at $\mathcal{O}(p^5)$, where a further expansion in the powers of $M$ of the chiral loops is performed, will change the analytical structures of the original amplitudes. This only brings minor effects for small pion masses. However, the difference becomes dramatic as the pion mass increases.
Specifically, as can be seen in Figure~\ref{fig:chiral.exp.mn.compare}, at the physical point, this operation leads to a difference $\sim 10$~MeV; however, it causes a dramatic deviation for large pion masses beyond $260$~MeV. The truncated IR prescription drops further terms that stem from the infrared regular parts of the loop integrals, and the resulting deviation is more sizable.

\section{Summary and conclusion\label{sec.summary}}

We have presented a leading two-loop analysis of the nucleon mass within the framework of covariant ChPT. The feasibility of the EOMS renormalization scheme at two-loop order is justified and demonstrated in detail. We address the notable PCB issue by using the dimensional counting analysis approach and can obtain an analytical expression of the chiral series up to any desired order. The full renormalized EOMS expression is considered to be the central result in this study, and additional truncations after further expansion in powers of $M$ either in the EOMS or IR approaches are worked out as well. 

The full EOMS result is shown to be suitable for performing chiral analysis of modern lattice QCD data with high precision. We find that the full EOMS result, possessing correct power counting and respecting the analytic property, can well describe the recent lattice QCD results~\cite{Hu:2024mas} in a broad range of pion masses. It is also found that the $\mathcal{O}(p^5)$ contribution from the full EOMS is small around $10$~MeV, implying that the chiral series in the full EOMS scheme converges very well. Nevertheless, we find that the truncated results after further expanding in powers of $M$ of the chiral loops are clearly different from the full EOMS case. Especially, the parameter $e_m$ from the two situations even differs in sign. Since the $\pi N$ sigma term $\sigma_{\pi N}$ is sensitive to $e_m$, an extension of the present work on the two-loop calculation of $\sigma_{\pi N}$ will definitely be helpful to address the discrepancy. 

\acknowledgments
We are grateful to Prof. Jambul~Gegelia for fruitful discussions on the EOMS renormalization at two-loop order. We would like to thank Prof. Yan-Qing~Ma, Prof. Jian~Wang and Prof. Yi-Bo~Yang for helpful discussions. DLY appreciates the support of Peng Huan-Wu visiting professorship and the hospitality of Institute of Theoretical Physics at Chinese Academy of Sciences (CAS), where part of this work was done. This work is supported by National Nature Science Foundations of China (NSFC) under Contract No.~12275076, No.~11905258, No.~12335002, No.~12047503, No.~12475078, and No.~12125507; by the Science Fund for Distinguished Young Scholars of Hunan Province under Grant No.~2024JJ2007; by the Fundamental Research Funds for the Central Universities under Contract No.~531118010379; by the Science Foundation of Hebei Normal University with Contract No. L2025B09 and No. L2023B09; by Science Research Project of Hebei Education Department under Contract No. QN2025063; and by the Chinese Academy of Sciences under Grant No. YSBR-101.

\appendix

\section{Definition of loop integrals}\label{sec.appA}

One-loop integrals are defined by 
\begin{align}\label{eq.def.olo.int}
J_{\nu_1\nu_2}^d=\int \frac{{\rm d}^d \ell}{{ i}\pi^{d/2}}\frac{1}{[\ell^2-M^2+i0^+]^{\nu_1}[(\ell+p)^2-m^2+i0^+]^{\nu_2}}\ , \quad (p^2=m_N^2)\ , 
\end{align}
with $d=4-2\epsilon$. For two-loop integrals, there are only $5$ independent irreducible scalar products that can be constructed from $\ell_1$, $\ell_2$ and $p$. Therefore, a convenient definition of two-loop integrals is given by
\begin{align}
I_{\nu_1\nu_2\nu_3\nu_4\nu_5}^d=
\int \frac{{\rm d}^d \ell_1}{{i}\pi^{d/2}}\frac{{\rm d}^d \ell_2}{{i}\pi^{d/2}} \frac{1}{\mathcal{D}_1^{\nu_1}\mathcal{D}_2^{\nu_2}\mathcal{D}_3^{\nu_3}\mathcal{D}_4^{\nu_4}\mathcal{D}_5^{\nu_5}}\ , \label{eq.int.tlo.def}
\end{align}
with denominators defined by
\begin{align}
&\mathcal{D}_1=\ell_1^2-M^2+i0^+ \ , \quad
\mathcal{D}_2=\ell_2^2-M^2+i0^+ \ , \quad 
\mathcal{D}_3=(\ell_1+\ell_2+p)^2-m^2+i0^+ \ , \notag \\
&\mathcal{D}_4=(\ell_1+p)^2-m^2+i0^+     \ , \quad
\mathcal{D}_5=(\ell_2+p)^2-m^2+i0^+ \ .
\end{align}

In dimensional regularization, a loop corresponds to an integration over the internal momentum like ${{\rm d}^d\ell}/{(2\pi)^d}$. Therefore, the above definitions of loop integrals imply that there actually exists a prefactor $\kappa$ for each loop:
\begin{align}
   \kappa = \frac{i}{16\pi^2} (4\pi \mu^2)^\epsilon \equiv \frac{i}{\Lambda} (4\pi \mu^2)^\epsilon \ ,\quad \Lambda =16\pi^2\ ,\quad \epsilon = \frac{4-d}{2}\ .\label{eq.kappa}
\end{align}
Here $\mu$ is an arbitrary parameter of mass dimension one. The introduction of $\mu$ is to preserve the dimensions of the loop integrals unchanged in $d$-dimensional spacetime.
In ChPT, the $\widetilde{{\rm MS}}=\overline{\rm MS}-1$ subtraction scheme is adopted, which can be imposed by making the substitution 
\begin{align}
  \mu^2\to \frac{\mu^2}{4\pi}
e^{
(\gamma_E-1)}\ .
\end{align}

\section{Rules for reduction to master integrals\label{sec.reduction.rule}}

For completeness, in this appendix, the reduction rules for the two-loop integrals in Section~\ref{sec.unren.mn} are explicitly shown below. There are in total $23$ integrals, which can be reduced by the IBP procedure~\cite{Smirnov:2019qkx}. The reduction relations we need are
\begin{align}
I_{00120}&=\frac{d-2}{2m^2}I_{00110} \ ,\\
I_{01(-1)20}&=\frac{(d-2)M^2-2m^2}{2m^2}I_{01010} \ ,\\
I_{01020}&=\frac{d-2}{2m^2}I_{01010} \ ,\\
I_{01120}&=-\frac{1}{2M^2 m^2(M^2-4m^2)}
\bigg\{ 
(d-2)\bigg[
M^2 I_{00110}-(M^2-2m^2)I_{01010}
\notag\\
&\hspace{4.2cm}-2m^2 I_{01100}
\bigg]
+2(d-3)M^2 m^2 I_{01110}
\bigg\}\ ,\\
I_{10120}&=\frac{1}{2M^2 m^2(M^2-4m^2)}
\bigg\{ 
2m^2\left[ 
(d-2)I_{10100}-(d-3)M^2 I_{10110}
\right]
-(d-2)M^2 I_{00110}
\bigg\}\ , \\
I_{11(-1)01}&=I_{11000}+M^2 I_{11001} \ , \\
I_{11(-1)10}&=I_{11000}+M^2 I_{11010} \ ,\\
I_{11(-1)11}&=\frac{1}{2m^2}\bigg\{ 
I_{00011}-I_{01010}-I_{10001}+I_{11000}
+M^2(I_{01011}-I_{10011})
\notag\\
&
-(M^2-2m^2)(I_{11001}+I_{11010})
+M^4 I_{11011}
\bigg\} \ ,\\
I_{11(-1)20}&=\frac{1}{2m^2(4m^2-M^2)}
\bigg\{
2m^2\bigg[ 
((d-4)M^2+4m^2)I_{11010}-(d-2)I_{11000}
\bigg]
+(d-2)M^2I_{01010}
\bigg\}\ , \\
I_{110(-1)0}&=M^2I_{11000} \ , \\
I_{110(-1)1}&=M^2I_{11001} \ , \\
I_{1100(-1)}&=M^2I_{11000} \ , \\
I_{1101(-1)}&=M^2I_{11010} \ .\\
I_{11020}&=\frac{1}{2M^2m^2(M^2-4m^2)}
\bigg\{ 
2m^2\left[(d-2)I_{11000}-(d-3)M^2I_{11010}\right]
-(d-2)M^2 I_{01010} 
\bigg\}
\ ,\\
I_{111(-1)(-1)}&=\frac{1}{3(3d-4)}\bigg\{
2\left[(4d-5)M^2-2(d-1)m^2\right](I_{01100}+ I_{10100})
\notag\\
&
+\left[(d-2)M^2+(d-1)m^2\right]
\big[ 2I_{11000}
-\frac{8M^2(M^2-m^2)}{d-2}(I_{12100}+I_{21100})
\big]
\notag \\
&+\frac{1}{d-2}
\bigg[ 
[(d-2)(13d-24)M^4-4(d-1)(7d-16)M^2 m^2
\notag \\
&
+8(d-1)(2d-5)m^4]I_{11100}
\bigg]
\bigg\} \ ,\\
I_{111(-1)0}
&=I_{1110(-1)}
=
\frac{1}{3(d-2)}
\bigg\{ 
4M^2(m^2-M^2)(I_{12100}+I_{21100})
\notag\\
&
+[(5d-12)M^2+2(5-2d)m^2]I_{11100}
\notag\\
&
+(d-2)(I_{01100}+I_{10100}+I_{11000})
\bigg\} \ ,\\
I_{1111(-1)}&=\frac{1}{2(d-2)m^2}
\bigg\{
(d-2)[M^4I_{11110}+M^2(I_{01110}+I_{10110})+I_{00110}-I_{01010}]
\notag\\
&
-(d-2)(M^2-2m^2)I_{11010}
+[(8-3d)M^2+2(d-3)m^2]I_{11100}
\notag\\
&
+4M^2(M^2-m^2)(I_{12100}+I_{21100})
\bigg\} \ ,\\
I_{111(-1)1}&= I_{1111-1} (\ell_1\leftrightarrow\ell_2) \ ,\\
I_{11120}&=-\frac{1}{2M^2 m^2(M^2-4m^2)}
\bigg\{
d M^4 I_{11110}+(d-2) M^2 (I_{01110}+ I_{10110})\notag\\
&-d M^2 I_{11010}-3 d M^2 I_{11100}+2 d m^2 I_{11010}-2 M^4 (I_{11110}-2I_{12100}-2I_{21100})\notag\\
&-4 M^2 m^2 (I_{11110}+ I_{12100}+ I_{21100})+2 M^2 (I_{11010}+4 I_{11100})-4 m^2 I_{11010}
\bigg\}\ ,
\\
I_{1200(-1)}&=\frac{d}{2}I_{11000} \ ,\\
I_{12000}&=\frac{d-2}{2M^2}I_{11000}\ ,\\
I_{1200(-1)}&=
\frac{1}{2M^2(M^2-4m^2)}
\bigg[ 
2(d-3)(M^2-2m^2)I_{11001}
+(d-2)(2I_{10001}-I_{11000})
\bigg\} \ .
\end{align}

\section{UV divergences of master integrals\label{sec.UV.div}}
In this Appendix, the UV divergences of the master integrals of one and two loops are shown explicitly. For convenience, we introduce two functions, $A_0(m_1^2)$ and $B_0(m_1^2, m_2^2; s)$, for the one-point and two-point one-loop scalar integrals, respectively. Their explicit expressions up to $\mathcal{O}(\epsilon)$ are 
\begin{align}
A_0(m_1^2)&=\frac{1}{\epsilon} m_1^2+m_1^2(1-\gamma_E-\log[m_1^2])
+\epsilon A_{\epsilon} \ , \label{eq.A0.eps.1}\\
B_0(m_1^2, m_2^2; s)&=\frac{1}{\epsilon}-B_{\rm fin.}(m_1^2,m_2^2;s)
+\epsilon B_{\epsilon} \ , 
\end{align}
where 
\bea
B_{\rm fin.}(m_1^2,m_2^2;s)=\int_0^1{\rm d}x
\left[\log\left(x m_1^2+(1-x)m_2^2-x(1-x)s\right)+\gamma_E\right] .
\eea
Furthermore, $A_{\epsilon}$ and $B_{\epsilon}$ are given by~\cite{Martin:2005qm}
\begin{align}
A_{\epsilon}&=m_1^2\big[ 
1-\frac{1}{2}(2-\gamma_E)\gamma_E+\frac12\zeta[2]-\log[m_1^2](1-\gamma_E-\frac12 \log[m_1^2]
\big] \ , \notag\\
B_{\epsilon}&=\frac12 \gamma_E^2+\frac12\zeta[2]+
\frac12\int^1_0 {\rm d}x
\log[x m_1^2+(1-x)m_2^2-x(1-x)s]\notag\\
&\hspace{4cm}\times(2\gamma_E+\log[x m_1^2+(1-x)m_2^2-x(1-x)s])
\ ,
\end{align}
with $\zeta$ being the Riemann zeta function.

At one-loop order, there are $3$ master integrals with the following UV divergences:
\begin{align}
J_{10}^{\rm div.}&=A_0^{\rm div.}(M^2) \ , 
\notag\\
J_{01}^{\rm div.}&=A_0^{\rm div.}(m^2) \ ,
\notag\\
J_{11}^{\rm div.}&=B_0^{\rm div.}(M^2, m^2; s) \ ,
\end{align}
where the superscript ${\rm div.}$ stands for the divergent terms proportional to $\epsilon^{-1}$.

There are 26 master integrals involved in the chiral corrections of two-loop accuracy to the nucleon mass. Note that only 10 of them are independent; see Eq.~\eqref{eq.tlo.master.int} and the discussion in Section~\ref{sec.unren.mn}. For master integrals with two different propagators, the UV divergent pieces read
\begin{align}
I^{\rm div.}_{00101}&=I^{\rm div.}_{00110}=\frac{m^4}{\epsilon^2}+\frac{2 m^4}{\epsilon}\left[1-\gamma_E-\log[m^2]\right] , \notag\\
I^{\rm div.}_{01100}&=
I^{\rm div.}_{10100}=
\frac{m^2 M^2}{\epsilon^2}+\frac{m^2 M^2}{\epsilon}\left[2(1-\gamma_E)-\log[m^2]-\log[M^2]\right] .
\end{align}
For master integrals with three different propagators, after defining
\begin{align}
S^{\rm div.}(m_1^2,m_2^2,m_3^2;s)=&
  \frac{1}{2\epsilon^2}(m_1^2+m_2^2+m_3^2)+
  \frac{1}{4\epsilon}\big[2(3-2\gamma_E)(m_1^2+m_2^2+m_3^2)\notag\\
  &
  -4(m_1^2\log m_1^2+m_2^2\log m_2^2+m_3^2\log m_3^2)-s\big] \ ,
\end{align}
the divergent parts can be written as 
\begin{align}
 I_{11100}^{\rm div.}&= S^{\rm div.}(M^2,M^2,m^2;m^2)\ ,\notag\\
  I_{01110}^{\rm div.}&= S^{\rm div.}(m^2,M^2,m^2;0)\ ,\notag\\
  I_{00111}^{\rm div.}&= S^{\rm div.}(m^2,m^2,m^2;m^2)\ ,\notag\\
  I_{10101}^{\rm div.}&= S^{\rm div.}(M^2,m^2,m^2;0)\ ,\notag\\
I_{12100}^{\rm div.}&=I_{21100}^{\rm div.}=\frac{1}{2\epsilon^2}+(\frac12-\gamma_E-\log[M^2])\frac{1}{\epsilon} \ ,\notag\\
I_{01101}^{\rm div.}&=
I_{10110}^{\rm div.}
=\frac{1}{\epsilon^2} m^2
+\frac{1}{\epsilon} m^2
\big[ 
1-\gamma_E-\log[m^2]
-B_{\rm fin.}(M^2,m^2;s)
\big] \ .
\end{align}
For master integrals with four different propagators, one has
\begin{align}
I_{11110}^{\rm div.}&=I_{11101}^{\rm div.}=U^{\rm div.}(M^2,m^2,M^2,m^2;m^2)\ ,\notag \\
I_{10111}^{\rm div.}&=I_{01111}^{\rm div.}=U^{\rm div.}(m^2,M^2,m^2,m^2;m^2)\ ,
\end{align}
with
\begin{align}
U^{\rm div.}(m_1^2,m_2^2,m_3^2,m_4^2;s)
=
\frac{1}{2\epsilon^2}
+\frac{1}{\epsilon}
\big[\frac{1}{2}-B_{\rm fin.}(m_1^2,m_2^2;s)\big]\ .
\end{align}

The others are reducible in the sense that they can be expressed as products of one-loop integrals, which read
\begin{align}
I_{00011}^{\rm div.}&=\left[A_0^{\rm div.}(m^2)\right]^2
=\frac{m^4}{\epsilon^2}+\frac{2m^4}{\epsilon}\left[1-\gamma_E-\log[m^2]\right] , \notag\\
I_{11000}^{\rm div.}&=\left[A_0^{\rm div.}(M^2)\right]^2
=\frac{M^4}{\epsilon^2}+\frac{2M^4}{\epsilon}\left[1-\gamma_E-\log[M^2]\right] , \notag \\
I_{10001}^{\rm div.}&=I_{01010}^{\rm div.}
=A_0^{\rm div.}(m^2) A_0^{\rm div.}(M^2)\notag\\
&=\frac{M^2 m^2}{\epsilon^2}+\frac{M^2 m^2}{\epsilon}\left[2(1-\gamma_E)-\log[M^2]-\log[m^2]\right] ,\\
I_{01011}^{\rm div.}&=I_{10011}^{\rm div.}=A_0^{\rm div.}(m^2) B_0^{\rm div.}(M^2, m^2; s)
\notag\\
&=\frac{1}{\epsilon^2}m^2+\frac{1}{\epsilon}m^2\left[ 
1-\gamma_E-\log[m^2]-B_{\rm fin.}(M^2, m^2;s) \right] ,
\notag\\
I_{11001}^{\rm div.}&=I_{11010}^{\rm div.}=A_0^{\rm div.}(M^2) B_0^{\rm div.}(M^2, m^2; s)
\notag\\
&=\frac{1}{\epsilon^2}M^2+\frac{1}{\epsilon}M^2
\left[1-\gamma_E-\log[M^2]-B_{\rm fin.}(M^2, m^2;s)\right] ,\\
I_{11011}^{\rm div.}&=B_0^{\rm div.}(M^2, m^2; s)B_0^{\rm div.}(M^2, m^2; s)
\notag\\
&=\frac{1}{\epsilon^2}
-\frac{2}{\epsilon}
B_{\rm fin.}(M^2, m^2;s) \ .
\end{align}

\section{Chiral expansion of master integrals\label{sec.dim.counting.analysis}}
The chiral expansion of all the master integrals can be conducted by employing the so-called dimensional counting analysis~\cite{Gegelia:1999qt} or strategy of regions~\cite{Smirnov:2002pj}. For one-loop master integrals, we have
\begin{align}\label{eq.ce.olo.master.int}
    J_{\nu_1\nu_2} = J_{\nu_1\nu_2}^{(0)} + J_{\nu_1\nu_2}^{(1)} 
\end{align}
where the infrared regular and singular parts, $J_{\nu_1\nu_2}^{(0)}$ and $J_{\nu_1\nu_2}^{(1)} $, are given by
\begin{align}
J_{\nu_1\nu_2}^{(0)} &= \sum_{i_1=0}^\infty \binom{-\nu_1}{i_1}(-1)^{i_1}M^{2i_1} \mathcal{R}_{\nu_1\nu_2}^{(i_1)}\ ,
\\
 J_{\nu_1\nu_2}^{(1)} &=\sum_{i_2=0}^\infty\binom{-\nu_2}{i_2}\bigg(\frac{M}{m}\bigg)^{d-2\nu_1-\nu_2+i_2}\mathcal{I}_{\nu_1\nu_2}^{(i_2)}\ ,\quad
\end{align}
with the binomial coefficient~\cite{Gradshteyn2014}
\begin{align}
\binom{m}{n}=\frac{\Gamma(n+1)}{\Gamma(m+1)\Gamma(n-m+1)}\ .
\end{align}
The coefficients $\mathcal{R}_{\nu_1\nu_2}^{(i_1)}$ and $\mathcal{I}_{\nu_1\nu_2}^{(i_2)} $ are independent of $M$, which are obtainable via
\begin{align}
\mathcal{R}_{\nu_1\nu_2}^{(i_1)} &=\int\frac{{\rm d}^d\ell}{i\pi^{d/2}}
\frac{1}{[\ell^2]^{\nu_1+i_1}[(\ell+p)^2-m^2]^{\nu_2}}\ ,\label{eq.R11.olo}\\
\mathcal{I}_{\nu_1\nu_2}^{(i_2)} &=\int\frac{{\rm d}^d\tilde{\ell}}{i\pi^{d/2}}
\frac{[\tilde{\ell}^2]^{i_2}}{[\tilde{\ell}^2-m^2]^{\nu_1}[2\tilde{\ell}\cdot p]^{\nu_2+i_2}}\ .\label{eq.I11.olo}
\end{align}
It is worth noting that additional UV divergences arise from $\mathcal{R}_{\nu_1\nu_2}^{(i_1)}$ and $\mathcal{I}_{\nu_1\nu_2}^{(i_1)}$, as pointed out by Ref.~\cite{Schindler:2007dr}. However, in Eq.~\eqref{eq.ce.olo.master.int}, the sum of those additional divergences vanishes.

For two-loop master integrals, the explicit form of chiral expansion reads 
\begin{align}\label{eq.ce.tlo.master.int}
I_{\nu_1\nu_2\nu_3\nu_4\nu_5}=I_{\nu_1\nu_2\nu_3\nu_4\nu_5}^{(0,0)}+I_{\nu_1\nu_2\nu_3\nu_4\nu_5}^{(1,0)}+I_{\nu_1\nu_2\nu_3\nu_4\nu_5}^{(0,1)}+I_{\nu_1\nu_2\nu_3\nu_4\nu_5}^{(1,1)}\ .
\end{align}
The overall factors in $M$ of the terms in the above decomposition are different, therefore they separately satisfy the Ward identities~\cite{Schindler:2003je}. 
The first term is called the infrared regular part, corresponding to the region where both integration momenta are hard. Its chiral expansion series reads
\begin{align}
I_{\nu_1\nu_2\nu_3\nu_4\nu_5}^{(0,0)}=
\sum_{i_1=0}^{\infty}\sum_{i_2=0}^{\infty}\binom{-\nu_1}{i_1}\binom{-\nu_2}{i_2}(-1)^{i_1+i_2} M^{2i_1+2i_2}\mathcal{R}^{(i_1,i_2)}_{\nu_1\nu_2\nu_3\nu_4\nu_5}\ ,
\end{align}
where the coefficients are given by
\begin{align}
\mathcal{R}^{(i_1,i_2)}_{\nu_1\nu_2\nu_3\nu_4\nu_5}  =\int \frac{{\rm d}^d\ell_1}{i\pi^{d/2}}\frac{{\rm d}^d\ell_1}{i\pi^{d/2}}\frac{1}{[\ell_1^2]^{\nu_1+i_1}}\frac{1}{[\ell_2^2]^{\nu_2+i_2}}\frac{1}{\mathcal{D}_3^{\nu_3}}\frac{1}{\mathcal{D}_4^{\nu_4}}\frac{1}{\mathcal{D}_5^{\nu_5}}\ .\label{eq.reg.tlo}
\end{align}
The second term of Eq.\eqref{eq.ce.tlo.master.int} is the infrared regular-singular mixing part. The corresponding chiral expression is 
\begin{align}
I_{\nu_1\nu_2\nu_3\nu_4\nu_5}^{(1,0)}&=  
\sum_{i_2=0}^{\infty}\sum_{i_3=0}^{\infty}\sum_{i_4=0}^{\infty}\sum_{k=0}^{i_3}
\binom{-\nu_2}{i_2}\binom{-\nu_3}{i_3}\binom{-\nu_4}{i_4}\binom{i_3}{k}(-1)^{i_2}M^{d-2\nu_1-\nu_4+2i_2+i_3+i_4+k}\notag \\
&\times m^{-d+2\nu_1+\nu_4-i_3-i_4-k} \mathcal{M}_{\nu_1\nu_2\nu_3\nu_4\nu_5}^{(i_2,i_3,i_4,k)}\ ,
\end{align}
with the coefficients
\begin{align}
\mathcal{M}_{\nu_1\nu_2\nu_3\nu_4\nu_5}^{(i_2,i_3,i_4,k)}
=\int\frac{{\rm d}^d\tilde{\ell}_1}{i\pi^{d/2}}\frac{{\rm d}^d\ell_2}{i\pi^{d/2}}
\frac{1}{[\tilde{\ell}_1^2-m^2]^{\nu_1}}\frac{1}{[\ell_2^2]^{\nu_2+i_2}}\frac{[2\tilde{\ell}_1\cdot(\ell_2+p)]^{i_3-k}}{[\ell_2^2+2\ell_2\cdot p]^{\nu_3+\nu_5+i_3}}
\frac{[\tilde{\ell}_1^2]^{i_4+k}}{[2\tilde{\ell}_1\cdot p]^{\nu_4+i_4}}
\ .\label{eq.mixing1.tlo}
\end{align}
For $I_{\nu_1\nu_2\nu_3\nu_4\nu_5}^{(0,1)}$, one only needs to interchange $\ell_1$ and $\ell_2$, and the explicit form is
\begin{align}
I_{\nu_1\nu_2\nu_3\nu_4\nu_5}^{(0,1)}&=  
\sum_{i_1=0}^{\infty}\sum_{i_3=0}^{\infty}\sum_{i_5=0}^{\infty}\sum_{k=0}^{\infty}
\binom{-\nu_1}{i_1}\binom{-\nu_3}{i_3}\binom{-\nu_5}{i_5}\binom{i_3}{k}(-1)^{i_1}
M^{d-2\nu_2-\nu_5+2i_1+i_3+i_5+k}\notag \\
&\times m^{-d+2\nu_2+\nu_5-i_3-i_5-k} 
\mathcal{M}_{\nu_1\nu_2\nu_3\nu_4\nu_5}^{(i_1,i_3,i_5,k)}\ ,
\end{align}
with the coefficients
\begin{align}
\mathcal{M}_{\nu_1\nu_2\nu_3\nu_4\nu_5}^{(i_1,i_3,i_5,k)}
=\int\frac{{\rm d}^d{\ell}_1}{i\pi^{d/2}}\frac{{\rm d}^d\tilde{\ell}_2}{i\pi^{d/2}}
\frac{1}{[{\ell}_1^2]^{\nu_1+i_1}}\frac{1}{[\tilde{\ell}_2^2-m^2]^{\nu_2}}\frac{[2\tilde{\ell}_2\cdot(\ell_1+p)]^{i_3-k}}{[\ell_1^2+2\ell_1\cdot p]^{\nu_3+\nu_4+i_3}}
\frac{[\tilde{\ell}_2^2]^{i_5+k}}{[2\tilde{\ell}_2\cdot p]^{\nu_5+i_5}}
\ .\label{eq.mixing2.tlo}
\end{align}
The last term of Eq.~\eqref{eq.ce.tlo.master.int} corresponds to the infrared singular part, which reads
\begin{align}
I^{(1,1)}_{\nu_1\nu_2\nu_3\nu_4\nu_5}
=\sum_{i_3=0}^{\infty}\sum_{i_4=0}^{\infty}\sum_{i_5=0}^{\infty}\binom{-\nu_3}{i_3}\binom{-\nu_4}{i_4}\binom{-\nu_5}{i_5}
\left(\frac{M}{m}\right)^{2d-2\nu_1-2\nu_2-\nu_3-\nu_4-\nu_5+i_3+i_4+i_5}
\mathcal{I}^{(i_3,i_4,i_5)}_{\nu_1\nu_2\nu_3\nu_4\nu_5} \ ,
\end{align}
with the coefficient 
\begin{align}
\mathcal{I}^{(i_3,i_4,i_5)}_{\nu_1\nu_2\nu_3\nu_4\nu_5} 
=\int\frac{{\rm d}^d\tilde{\ell}_1}{i\pi^{d/2}}
\frac{{\rm d}^d\tilde{\ell}_2}{i\pi^{d/2}}
\frac{1}{[\tilde{\ell}_1^2-m^2]^{\nu_1}}
\frac{1}{[\tilde{\ell}_2^2-m^2]^{\nu_2}}
\frac{[\tilde{\ell}_1^2+\tilde{\ell}_2^2+2\tilde{\ell}_1\cdot\tilde{\ell}_2]^{i_3}}{[2(\tilde{\ell}_1+\tilde{\ell}_2)\cdot p]^{\nu_3+i_3}}
\frac{[\tilde{\ell}_1^2]^{i_4}}{[2\tilde{\ell}_1\cdot p]^{\nu_4+i_4}}
\frac{[\tilde{\ell}_2^2]^{i_5}}{[2\tilde{\ell}_2\cdot p]^{\nu_5+i_5}} \ .\label{eq.sig.tlo}
\end{align}
The infrared singular part does not violate power counting, and the IR prescription keeps this part only.

\section{Explicit expressions of basic integrals in chiral expansion\label{sec.exp.master.int}}

In Appendix~\ref{sec.dim.counting.analysis}, the chiral expansions of the master integrals introduce the following integral coefficients: $\mathcal{R}_{\nu_1\nu_2}^{(i_1)}$~\eqref{eq.R11.olo}, $\mathcal{I}_{\nu_1\nu_2}^{(i_1)}$~\eqref{eq.I11.olo}, $\mathcal{R}^{(i_1,i_2)}_{\nu_1\nu_2\nu_3\nu_4\nu_5}$~\eqref{eq.reg.tlo}, $\mathcal{M}_{\nu_1\nu_2\nu_3\nu_4\nu_5}^{(i_2,i_3,i_4,k)}$~\eqref{eq.mixing1.tlo}, $\mathcal{M}_{\nu_1\nu_2\nu_3\nu_4\nu_5}^{(i_1,i_3,i_5,k)}$~\eqref{eq.mixing2.tlo} and $\mathcal{I}^{(i_3,i_4,i_5)}_{\nu_1\nu_2\nu_3\nu_4\nu_5}$~\eqref{eq.sig.tlo}. Analogously to the one- and two-loop integrals defined in Appendix~\ref{sec.appA}, those integral coefficients can be reduced by PV or IBP procedures and, finally, are expressed in terms of a set of basic integrals. In this appendix, we collect all the basic integrals needed in our calculation and derive analytic expressions for them. 

For later convenience, we introduce the analytic form of one-point one-loop function,
\begin{align}
    A_0(m^2)=-\left(m^2\right)^{{d}/{2}-1} \Gamma \left(1-\frac{d}{2}\right)\ .
\end{align}
Note that Eq.~\eqref{eq.A0.eps.1} corresponds to the expansion of the above expression up to the linear term of $\epsilon$ only. For one-loop coefficients, four basic integrals are required:
\begin{align}
   \mathcal{R}^{\rm reg.}_{01}&\equiv \mathcal{R}_{01}^{(0)} = A_0(m^2) \ ,\label{eq.ono.exp.reg}\\
   \mathcal{R}^{\rm reg.}_{11}&\equiv \mathcal{R}_{11}^{(0)}= \frac{(d-2)}{2(d-3)m^2}A_0(m^2)\ ,\\
   \mathcal{I}^{\rm sig.}_{10}&\equiv \mathcal{I}_{10}^{(0)} =A_0(m^2)\ ,\\
\mathcal{I}^{\rm sig.}_{11}&\equiv\mathcal{I}_{11}^{(0)}=\frac{\sqrt{\pi }}{2}  \left(m^2\right)^{{(d-4)}/{2}} \Gamma \left(\frac{3}{2}-\frac{d}{2}\right)\ .
\end{align}

For two-loop regular-part coefficients, we define
\begin{align}\label{eq.reg.master.tlo}
\mathcal{R}_{\alpha_1\alpha_2\alpha_3\alpha_4\alpha_5}^{\rm reg}
 =\int \frac{{\rm d}^d\ell_1}{i\pi^{d/2}}\frac{{\rm d}^d\ell_1}{i\pi^{d/2}}\frac{1}{[D_1^{\rm reg}]^{\alpha_1}
 [D_2^{\rm reg}]^{\alpha_2}
 [D_3^{\rm reg}]^{\alpha_3}
 [D_4^{\rm reg}]^{\alpha_4}
 [D_5^{\rm reg}]^{\alpha_5}
 }\ ,
\end{align}
with the denominators
\begin{align}
D_1^{\rm reg} = \ell_1^2\ ,\quad 
D_2^{\rm reg} = \ell_2^2\ , \quad
D_3^{\rm reg} = D_3\ ,\quad
D_4^{\rm reg} = D_4\ ,\quad
D_5^{\rm reg} = D_5\ .
\end{align}
The coefficients $\mathcal{R}^{(i_1,i_2)}_{\nu_1\nu_2\nu_3\nu_4\nu_5}$ \eqref{eq.reg.tlo} can be reexpressed in terms of the above integral family. Furthermore, all the integrals can be reduced to a linear combination of the following ones:
\begin{align}
\mathcal{R}_{00011}^{\rm reg}&=[A_0(m^2)]^2=(m^2)^{d-2}\left(\Gamma\left[1-\frac{d}{2}\right]\right)^2 \ ,\\
\mathcal{R}_{00101}^{\rm reg}&=[A_0(m^2)]^2=(m^2)^{d-2}\left(\Gamma\left[1-\frac{d}{2}\right]\right)^2 \ ,\\
\mathcal{R}_{00110}^{\rm reg}&=[A_0(m^2)]^2=(m^2)^{d-2}\left(\Gamma\left[1-\frac{d}{2}\right]\right)^2 \ ,\\
\mathcal{R}_{00111}^{\rm reg}&=
\frac{2^{-d-3} (d-2) \pi ^{\frac{3}{2}-d} (m^2)^{d-3} \csc \left(\frac{\pi  d}{2}\right) \Gamma(2-d)}{\Gamma \left(\frac{5}{2}-\frac{d}{2}\right) \Gamma \left(\frac{d}{2}\right)}\notag\\
&\times
\bigg[(d-3) \, _3F_2\left(1,\frac{4-d}{2},\frac{d-1}{2};\frac{3}{2},d-1;1\right)+\, _3F_2\left(\frac{1}{2},1,3-d;\frac{5-d}{2},\frac{d}{2};1\right)\bigg]
\ , \label{eq.reg.00111}\\
\mathcal{R}_{11100}^{\rm reg}&=-(m^2)^{d-3}\frac{\Gamma[3-d] \left(\Gamma\left[1-\frac{d}{2}\right]\right)^2 
\Gamma[2-\frac{d}{2}]}{\Gamma[\frac{d}{2}]}\ 
_2F_1\left(2-\frac{d}{2},3-d;\frac{d}{2};\frac{p^2}{m^2}\right)
\ .\label{eq.reg.11100}
\end{align}

Likewise, for the mixing-part coefficients, we define
\begin{align}
\mathcal{M}^{
\rm mix_{10}
}_{\beta_1\beta_2\beta_3\beta_4\beta_5}=
\int
\frac{{\rm d}^d\tilde{\ell}_1}{i\pi^{d/2}}
\frac{{\rm d}^d\ell_2}{i\pi^{d/2}}
\frac{
1
}{
[D_1^{\rm mix_{10}}]^{\beta_1}
[D_2^{\rm mix_{10}}]^{\beta_2}
[D_3^{\rm mix_{10}}]^{\beta_3}
[D_4^{\rm mix_{10}}]^{\beta_4}
[D_5^{\rm mix_{10}}]^{\beta_5}
}\ ,
\end{align}
with the denominators
\begin{align}
D_1^{\rm mix_{10}}=  \tilde{\ell}_1^2-m^2\ ,\,
D_2^{\rm mix_{10}}=  {\ell}_2^2\ ,\,
D_3^{\rm mix_{10}}= (\tilde{\ell}_1+{\ell}_2)^2\ ,\,
D_4^{\rm mix_{10}}= 2\tilde{\ell}_1\cdot p\ ,\,
D_5^{\rm mix_{10}}=D_5\ .
\end{align} 
The coefficients $\mathcal{M}_{\nu_1\nu_2\nu_3\nu_4\nu_5}^{(i_2,i_3,i_4,k)}$~\eqref{eq.mixing1.tlo} can be recast to the above integral family. The mixing-part integrals involved in our calculation can be reduced to the following two basic integrals
\begin{align}
\mathcal{M}^{
\rm mix_{10}}_{10001}&= \mathcal{I}_{10}^{(0)}\, \mathcal{R}_{01}^{(0)}=[A_0(m^2)]^2\ ,\\
\mathcal{M}^{\rm mix_{10}}_{10011}&=\mathcal{I}_{11}^{(0)}\, \mathcal{R}_{01}^{(0)}\ .
\end{align}
The coefficients $\mathcal{M}_{\nu_1\nu_2\nu_3\nu_4\nu_5}^{(i_1,i_3,i_5,k)}$~\eqref{eq.mixing2.tlo} can be treated in the same way as $\mathcal{M}_{\nu_1\nu_2\nu_3\nu_4\nu_5}^{(i_2,i_3,i_4,k)}$~\eqref{eq.mixing1.tlo}, by substituting $\tilde{\ell}_1\to\tilde{\ell}_2$ and $\ell_2\to\ell_1$. The basic integrals we need are
\begin{align}
\mathcal{M}^{\rm mix_{01}}_{01010}&=\mathcal{I}_{10}^{(0)}\, \mathcal{R}_{01}^{(0)}=[A_0(m^2)]^2\ ,\\
\mathcal{M}^{\rm mix_{01}}_{01011}&=\mathcal{I}_{11}^{(0)}\, \mathcal{R}_{01}^{(0)}\ .
\end{align}

For the singular-part coefficients, two families of integrals are required. The first family is defined by
\begin{align}
\mathcal{I}^{\rm sin
}_{\gamma_1\gamma_2\gamma_3\gamma_4\gamma_5}=
\int
\frac{{\rm d}^d\tilde{\ell}_1}{i\pi^{d/2}}
\frac{{\rm d}^d\tilde{\ell}_2}{i\pi^{d/2}}
\frac{1}{
[D_1^{\rm sin}]^{\gamma_1}
[D_2^{\rm sin}]^{\gamma_2}
[D_3^{\rm sin}]^{\gamma_3}
[D_4^{\rm sin}]^{\gamma_4}
[D_5^{\rm sin}]^{\gamma_5}
}\ ,
\end{align}
with the denominators
\begin{align}
D_1^{\rm sin} = \tilde{\ell}_1^2-m^2\ ,
D_2^{\rm sin} = \tilde{\ell}_2^2-m^2\ ,
D_3^{\rm sin} = (\tilde{\ell}_1+ \tilde{\ell}_2)^2\ ,
D_4^{\rm sin} = 2\tilde{\ell}_1\cdot p\ ,
D_5^{\rm sin} = 2(\tilde{\ell}_2+\tilde{\ell}_1)\cdot p\ .
\end{align}
The second family, denoted by $\mathcal{I}^{\rm sinp
}_{\gamma_1\gamma_2\gamma_3\gamma_4\gamma_5}$, is different from the first one by the last denominator, 
\begin{align}
D_1^{\rm sinp} = D_1^{\rm sin}\ ,
D_2^{\rm sinp} = D_2^{\rm sin}\ ,
D_3^{\rm sinp} = D_3^{\rm sin}\ ,
D_4^{\rm sinp} = D_4^{\rm sin}\ ,
D_5^{\rm sinp} = 2\tilde{\ell}_2\cdot p\ .
\end{align}
In the case that $D_5^{\rm sin}$ and $D_5^{\rm sinp}$ appear simultaneously in the same denominator, we separate them using the identity,
\begin{align}
\frac{1}{D_5^{\rm sin} D_5^{\rm sinp}}
=\frac{1}{D_4^{\rm sinp}D_5^{\rm sinp}}
-\frac{1}{D_4^{\rm sin}D_5^{\rm sin}}\ .
\end{align}
The coefficients $\mathcal{I}^{(i_3,i_4,i_5)}_{\nu_1\nu_2\nu_3\nu_4\nu_5}$~\eqref{eq.sig.tlo} are related to the two families, which can eventually be reduced to the basic integrals:
\begin{align}
\mathcal{I}^{\rm sin
}_{11000} & =[\mathcal{I}_{10}^{(0)}]^2=[A_0(m^2)]^2\ , \\
\mathcal{I}^{\rm sin
}_{11001} &=-\frac{\sqrt{\pi } \left({m}^2\right)^{d-3} \Gamma \left(\frac{5}{2}-d\right) \Gamma \left(\frac{3}{2}-\frac{d}{2}\right)^2}{2 \Gamma (3-d)}\ ,
\\
\mathcal{I}^{\rm sin
}_{11010} &=-2^{d-2} {m}^{2 d-6} \pi  \Gamma (2-d)\ ,\\
\mathcal{I}^{\rm sin}_{11011}&= \frac{1}{8} \pi  {m}^{2 d-8} \Gamma \left(\frac{3}{2}-\frac{d}{2}\right)^2\ ,\\
\mathcal{I}^{\rm sinp
}_{11000} &=[\mathcal{I}_{10}^{(0)}]^2=[A_0(m^2)]^2\ ,\\
\mathcal{I}^{\rm sinp
}_{11001} &=\mathcal{I}_{10}^{(0)}\mathcal{I}_{11}^{(0)}\ ,\\
\mathcal{I}^{\rm sinp
}_{11011} &=[\mathcal{I}_{11}^{(0)}]^2\ .
\end{align}

\section{One-loop renormalization factors\label{sec.helper}}
The one-loop wave function renormalization constant of the nucleon can be found in Ref.~\cite{Chen:2012nx,Siemens:2016hdi}, which reads
\begin{align}
\mathcal{Z}_N^\text{(1-loop)} = 1+ \delta\mathcal{Z}_N^{(2)}+\cdots\ ,\quad \label{eq.1loop.ZN}
\end{align}
where the $\mathcal{O}(p^2)$ correction, corresponding to diagram~(a) of Fig.~\ref{fig:1loop}, reads
\begin{align}
 \delta\mathcal{Z}_N^{(2)}&=-\frac{3g^2\,\kappa}{4F^2(M^2-4m^2)}\bigg[ 
2M^2\left[(d-2)M^2-2(d-1)m^2 \right] J_{11}
\notag \\
&+\left[(3-2d)M^2+4(d-1)m^2\right]J_{10}
+2(d-2)M^2 J_{01}
\bigg]   \ .
\end{align}
The physical quantities $g_A$, $M_\pi$ and $F_\pi$ are related to the renormalized ones through
\begin{align}
g_A &= g\bigg[1+\frac{4 {d}^r_{16} {M}^2}{g}+\frac{1}{\Lambda}\delta g^{(2)}_{\rm 1-loop} 
\bigg]\label{eq.gA.1loop}\ ,\\
M_\pi^2 &= M^2 \bigg[1+\frac{2{\ell}_3^r M^2}{F^2}-\frac{\bar{J}_{10}}{2F^2\Lambda}\bigg] , \\
F_\pi &= F\bigg[1+{\ell}_4^r\frac{M^2}{F^2}+
\frac{\bar{J}_{10}}{F^2\Lambda} \bigg] \ ,\label{eq.Fpi.1loop}
\end{align}
where the one-loop correction to the axial charge of the nucleon reads
\begin{align}
\delta g^{(2)}_\text{1-loop} & = \frac{g^2 m^2 \left(3 {M}^2+4 {m}^2\right)}{F^2 \left({M}^2-4 {m}^2\right)}
+\frac{\left(8 \left({g}^2+1\right) {M}^2 {m}^2-\left(3 {g}^2+2\right) {M}^4\right) \bar{J}_{11}}{F^2 \left({M}^2-4 {m}^2\right)}
\notag \\
&
+\frac{ \left(\left(4 {g}^2+1\right) {M}^2-4 \left(2 {g}^2+1\right) {m}^2\right)\bar{J}_{10}}{F^2 \left({M}^2-4 {m}^2\right)}-\frac{ \left(\left(3 {g}^2+2\right) {M}^2+4 \left({g}^2-2\right) {m}^2\right)\bar{J}_{01}}{F^2 \left({M}^2-4 {m}^2\right)}\ .
\end{align}

\bibliography{refs.bib}
\end{document}